\documentclass{article}

\PassOptionsToPackage{numbers, sort&compress}{natbib}

\usepackage[preprint]{neurips_2023}
\usepackage[utf8]{inputenc}
\usepackage[T1]{fontenc}
\usepackage{hyperref}
\usepackage{url}
\usepackage{booktabs}
\usepackage{amsfonts}
\usepackage{nicefrac}
\usepackage{microtype}
\usepackage{xcolor}
\usepackage[pdftex]{graphicx}
\usepackage{tabularx}
\usepackage{subfigure}
\usepackage{color,soul}
\title{BMW Agents - A Framework For Task Automation Through 
Multi-Agent Collaboration}

\author{
Noel Crawford \quad Edward B. Duffy \quad Iman Evazzade \quad Torsten Foehr \\
\quad \textbf{Gregory Robbins} \quad \textbf{Debbrata Kumar Saha} 
\quad \textbf{Jiya Varma}  \quad \textbf{Marcin Ziolkowski}$^\dag$ \thanks{Authors in alphabetical order.}\\ 
\\
BMW Group\\
Information Technology Research Center \\
2 Research Drive, Greenville, SC 29607 \\
\\
$^\dag$ marcin.ziolkowski@bmwgroup.com\\
}

\begin{document}

\maketitle

\begin{abstract}
Autonomous agents driven by Large Language 
Models (LLMs) offer enormous potential for automation. Early proof 
of this technology can be found in various demonstrations of agents solving 
complex tasks, interacting with external systems to augment their knowledge, 
and triggering actions. In particular, workflows involving multiple agents 
solving complex tasks in a collaborative fashion exemplify their capacity 
to operate in less strict and less well-defined environments. 
Thus, a multi-agent approach has great potential for serving as a backbone 
in many industrial applications, ranging from complex knowledge retrieval 
systems to next generation robotic process automation. Given the reasoning 
abilities within the current generation of LLMs, complex processes require 
a multi-step approach that includes a plan of well-defined and modular tasks. 
Depending on the level of complexity, these tasks can be executed either by 
a single agent or a group of agents. In this work, we focus on designing a
flexible agent engineering framework with careful attention to planning and execution,
capable of handling complex use case applications across various domains.
The proposed framework provides reliability in industrial applications and 
presents techniques to ensure a scalable, flexible, and collaborative 
workflow for multiple autonomous agents working together towards solving tasks.
\end{abstract}

\section{Introduction}\label{introduction}

In the rapidly evolving landscape of artificial intelligence (AI), the 
deployment of generative AI models marks a significant technological 
advancement, transforming how businesses and organizations recognize the value 
of AI and its potential for automation of complex tasks. While the emerging 
capabilities of Large Language Models (LLMs) are impressive, their applications 
in industrial settings are limited within this current generation of models. 
By themselves, LLMs do not have access to confidential and proprietary business
information which is necessary for developing robust and high quality AI-powered 
applications. While it is possible to fine-tune these models with 
company-specific data, one must carefully consider the challenges associated 
with this approach, such as data preparation and maintainability. Additionally, 
there is a large IT ecosystem where existing tools and applications can be 
leveraged without the need to duplicate data. These limitations highlight 
the need of a more dynamic approach to AI application design using AI agents 
\cite{park2023generative}. This work explores an approach that leverages the 
capabilities of AI agents to significantly enhance productivity and innovation 
through collaborative multi-agent workflows.

The integration of multiple AI agents in a cohesive workflow presents a
paradigm shift from traditional, singular AI applications to a more
dynamic, interconnected framework. This method not only amplifies the
individual capabilities of each agent, which can now have narrow
expertise and operate robustly, it also orchestrates a workflow of
interactions that drive complex task completion 
\cite{Besta_2024, xu2023rewoo}. This document will
detail the foundational principles of designing and implementing a
robust multi-agent engineering framework \cite{agent_engineering_book}. It will discuss the architectural
considerations, chosen algorithms, and methods for realizing complex
generative AI applications.

The trajectory of AI agents research, particularly with a focus on LLMs, 
took a notable turn with the advent of specialized frameworks aimed at the 
completion of very generic tasks. With the introduction of projects like 
AutoGPT \cite{autogpt} and BabyAGI \cite{babyagi}, the interest in AI agents expanded from academic research 
settings into a broader community. Both projects were able to demonstrate that 
complex tasks can be decomposed into simpler steps. By programmatically 
orchestrating the solution of each individual step, a solution to the more 
complex task can be achieved. In the scope of AI agents powered by LLMs, 
the ideas surrounding task decomposition, planning, and task execution using 
tools have emerged to overcome the limitations observed in the existing 
generation of LLMs \cite{xu2023rewoo}.

In this work we turn our attention to multi-agent systems 
\cite{wu2023autogen, liu2023dynamic, chen2023agentverse, guo2024embodied, 
chen2024llmarena, rasal2024llm, li2023camel, wang2024unleashing, 
fu2023improving} and the
potential of solving complex tasks through the collaborative work of AI
agents. Our perspective is rooted in the observations that 1) AI agents
perform well if their scope of responsibility is narrowed to a well
defined role \cite{li2024agents}, 2) complex problems require the expertise of several 
narrowly defined AI agents to successfully mimic human solutions with step-by-step 
reasoning and deliberation
\cite{wei2023chainofthought, xu2023expertprompting, abdelnabi2023llmdeliberation}, 
3) simple tasks can be completed with more 
complex prompting strategies like ReAct \cite{react2023}, and 4) impactful 
adoption of AI applications cannot be achieved in isolation from existing
development ecosystems \cite{tang2023toolalpaca, patil2023gorilla}. Based on these observations, we have developed an agent engineering framework to enable task
automation through multi-agent collaboration in enterprise settings. The framework is meant to interact within an IT landscape
that is a mixture of modern and legacy applications. It serves as a
template for stable AI applications that enable the automation of
complex processes, either as human or programmatically triggered
workflows. This framework is designed to be modular, extendable, and
stay agnostic to the dynamic LLM landscape.

This report serves as a technical introduction of methodology and an example blueprint 
for industrial applications of multi-agent workflows. It aims to offer valuable insights
and guidelines for organizations looking to leverage these advanced AI
systems to achieve scalability, resilience, and a competitive edge. It may also serve as 
a didactic resource for applied researchers new to the field of AI 
driven automation and AI agents. This document is organized in the following way: 
After the introduction,
we highlight the current AI agent landscape in Section \ref{existing-frameworks}. 
In Section \ref{agent-workflow} we present our agent workflow 
design and the definition of components. Section \ref{multi-agent-workflow} covers approaches for utilizing groups 
of AI agents, architecture of solution, and communication strategies. 
We focus on examples in Section \ref{example-applications} and
conclude with the lessons learned in Section \ref{summary}.

\section{Existing Frameworks}\label{existing-frameworks}

Table \ref{table-existing-work} includes a list of 
existing agent frameworks or projects closely related to 
agent workflows. The list is not all-inclusive as the landscape is rapidly growing.
Detailed discussions of existing approaches are covered in recently
published surveys \cite{wang2024survey, guo2024large, 
masterman2024landscape, xi2023rise, du2024survey}.

\begin{table}
  \caption{Sample of existing approaches to agent and multi-agent frameworks.
  Our view of their unique contributions is mentioned in the \textbf{Features} 
  column. Reference to project website and GitHub repository are included in the 
  \textbf{Project spaces} column if available.
  \label{table-existing-work}
  }
  \centering
  \begin{tabularx}{\linewidth}{XXX}
    \toprule
    \textbf{Name}     & \textbf{Features}     & \textbf{Project spaces} \\
    \midrule
AutoGen \cite{wu2023autogen} & Multi-agent, flexible
communication strategies &
\href{https://microsoft.github.io/autogen/}{Project} and
\href{https://github.com/microsoft/autogen}{GitHub} \\
\midrule
AutoGPT & Autonomous agents, dynamic agent creation, multiple LLMs &
\href{https://autogpt.net/}{Project} and
\href{https://github.com/Significant-Gravitas/AutoGPT}{GitHub} \\
\midrule
LangChain & Tool usage, interaction with existing applications &
\href{https://www.langchain.com}{Project} and
\href{https://github.com/langchain-ai/langchain}{GitHub} \\
\midrule
LangGraph & Multi-agent generalization of LangChain &
\href{https://python.langchain.com/docs/langgraph/}{Project} and
\href{https://github.com/langchain-ai/langgraph}{GitHub} \\
\midrule
LlamaIndex & Knowledge retrieval and RAG methods &
\href{https://www.llamaindex.ai/}{Project} and
\href{https://github.com/run-llama/llama_index}{GitHub} \\
\midrule
ChatDev \cite{qian2023communicative} &
Multi-agent, software development, agent dialog &
\href{https://chatdev.ai/}{Project} and
\href{https://github.com/OpenBMB/ChatDev}{GitHub} \\
\midrule
RAISE \cite{liu2024llm} & ReAct-like prompt strategy for multi-agent workflow & \\
\midrule
GPT-Engineer & Software development &
\href{https://github.com/gpt-engineer-org/gpt-engineer}{GitHub} \\
\midrule
MetaGPT \cite{hong2023metagpt} & Multi-agent, software
development & \href{https://www.deepwisdom.ai/}{Project} and
\href{https://github.com/geekan/MetaGPT}{GitHub} \\
\midrule
DyLAN \cite{liu2023dynamic} & Agent team optimization,
inference time agent selection &
\href{https://github.com/SALT-NLP/DyLAN}{GitHub} \\
\midrule
AgentVerse \cite{chen2023agentverse} & Dynamic agent group selection &
\href{https://agentverse.ai}{Project} and
\href{https://github.com/OpenBMB/AgentVerse}{GitHub} \\
\midrule
Embodied Agents \cite{guo2024embodied} & Agent team organization and 
communication patterns & \\
\midrule
AgentLite \cite{liu2024agentlite} & Multi-agent, hierarchical
agent orchestration, task decomposition &
\href{https://github.com/SalesforceAIResearch/AgentLite}{GitHub} \\
\midrule
LLMArena \cite{chen2024llmarena} & Multi-agent logic game playing& \\
\midrule
LLMHarmony \cite{rasal2024llm} & CAMEL \cite{li2023camel}
prompt extension to many agents &
\href{https://github.com/sumedhrasal/simulation}{GitHub} \\
\midrule
AgentGPT & No-code agent setup & 
\href{https://github.com/reworkd/AgentGPT}{GitHub} \\
\midrule
crewAI & Multi-agent with sequential and hierarchical strategy & \href{https://www.crewai.com/}{Project} and
\href{https://github.com/joaomdmoura/crewAI}{GitHub} \\
\midrule
SuperAGI & Pausing and resuming agents &
\href{https://github.com/TransformerOptimus/SuperAGI}{GitHub} \\
\midrule
BabyAGI & Reference concepts of AI agent workflow &
\href{https://github.com/yoheinakajima/babyagi}{GitHub} \\
\midrule
OpenAgents \cite{xie2023openagents} & Discussion of robustness of 
implementation, dynamic plugin selection &
\href{https://docs.xlang.ai/user-manual/overview}{Project} and
\href{https://github.com/xlang-ai/OpenAgents}{GitHub} \\
\midrule
HuggingFace \newline Transformers Agents & Long-memory,
multi-agent collaboration, rich ecosystem and active community &
\href{https://huggingface.co/blog/agents}{Project} \\
\bottomrule
  \end{tabularx}
\end{table}

\section{Agent Workflow}\label{agent-workflow}

Special attention is given to creating a flexible structure for agent workflows
which can support both single and multi-agent execution. The agent workflow in our implementation will follow the three main stages:

\begin{enumerate}
\def\labelenumi{\arabic{enumi}.}
\item
  \textbf{Planning} - Decomposition of the input into simple logical steps 
  with a clearly defined order of operations.
\item
  \textbf{Execution} - Completion of the work planned in Step (1) by agents
  solving simple tasks and creating results.
\item
  \textbf{Verification} - Independent check if the original objective has been
  achieved in step (2).
\end{enumerate}

\begin{figure}[ht]
\centering
\includegraphics[width=0.7\linewidth]{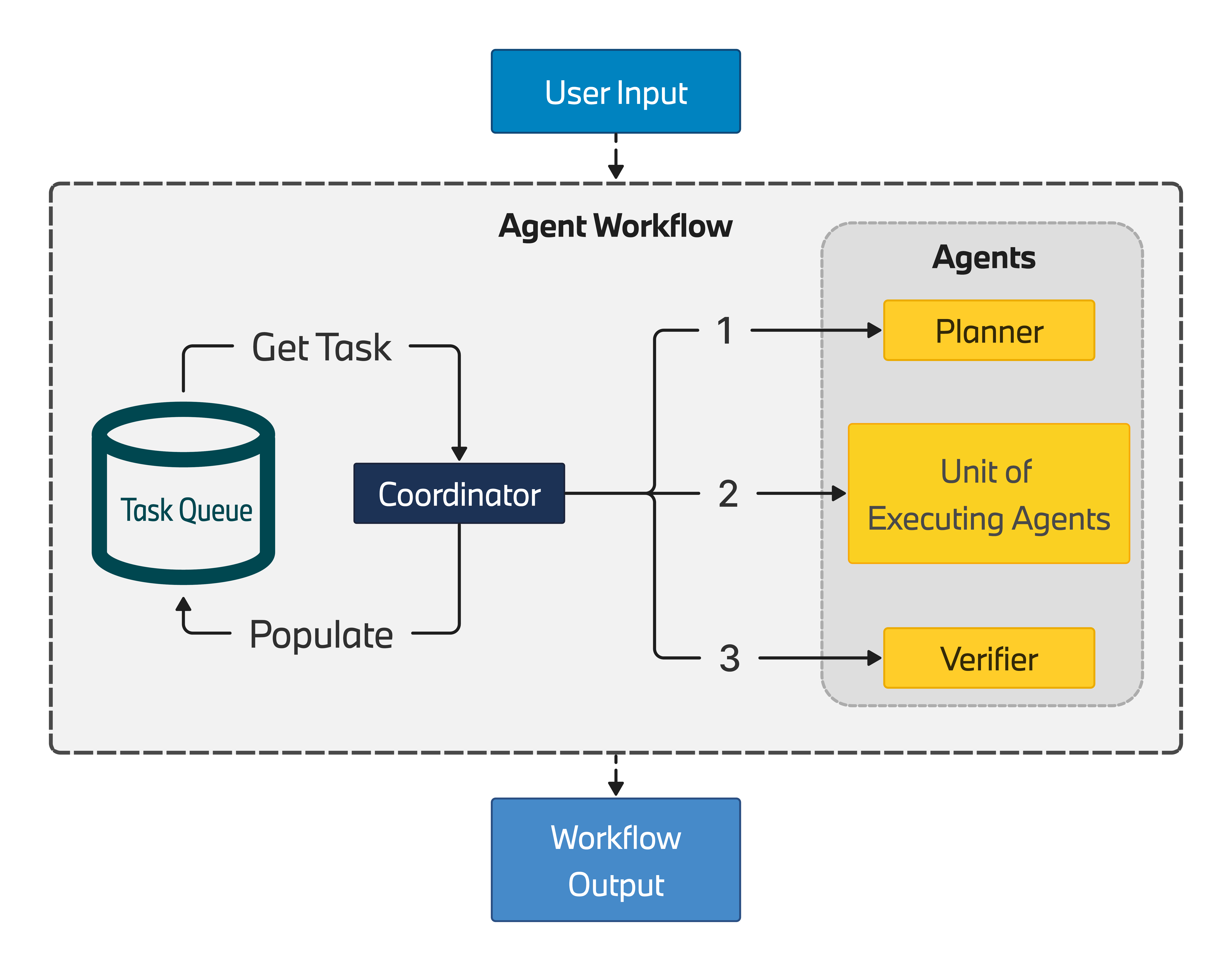}
\caption{Generic agent workflow starting with user input and ending with
providing workflow output. \textbf{Agent Workflow} highlights major components
and levels of the workflow with (1) Planning, (2) Execution, and (3)
Verification done by dedicated agents. \label{figure-generic-workflow}}
\end{figure}

The flow, which starts with user-provided (or in general, externally-provided) input is illustrated in Figure \ref{figure-generic-workflow}.

For readability, Figure \ref{figure-generic-workflow} includes a unit of 
executing agents, without a detailed composition of this unit,
directly connected to \textbf{Coordinator}. The above-mentioned note about 
``external input'' generalizes to agent workflows that may be triggered programmatically as opposed to human provided input.

The individual components involved in our generic agent workflow may be
defined as:

\begin{quote}
An \textbf{Agent} is an object that makes LLM calls with a
specified prompt strategy in order to achieve a specific result. Agent
includes a \emph{persona} which defines its scope of knowledge and
actions. Potentially, an Agent can have access to a set of tools, each capable of taking specific actions. Any available tool will
have a brief description along with instructions for the input schema included in the Agent's prompt.
\end{quote}

\begin{quote}
An \textbf{Agent Unit} is a container that holds one or more Agents that are meant to work together to solve a task. The Agent Unit is responsible for selecting an individual Agent in each iteration of solving a task. Selection strategies can range from simple patterns to more dynamic approaches using AI. Within a workflow, multiple Agent Units may be employed to facilitate intricate and specialized patterns of collaboration.
\end{quote}

\begin{quote}
A \textbf{Matcher} is an abstraction layer that is used to enable various methods of selecting an item based on some criteria. One potential use of a Matcher is selecting an appropriate Agent Unit for a given task. Matching can happen through simple methods such as list iteration or through more advanced methods involving AI.  
\end{quote}

\begin{quote}
The \textbf{Executor} is the component which orchestrates all Agent relevant 
operations through interactions with an Agent Unit. It is also responsible for bringing information related to the current running task to the Agent in the Unit.
\end{quote}

\begin{quote}
\textbf{Tools} are external functions that are made available to an Agent to complete a task. This can include access to databases, file systems, and APIs. Together, the set of Tools form the Agent's Toolbox.
\end{quote}

\begin{quote}
The \textbf{Toolbox Refiner} is responsible for filtering a set of Tools into a relevant subset during the execution of an Agent loop. This can improve the accuracy and performance of an Agent by limiting the number of Tool options. 
\end{quote}

\begin{quote}
The \textbf{Coordinator} orchestrates the agent workflow and executes
all components needed for the planning, execution, and verification of the data
flow.
\end{quote}

\begin{quote}
The \textbf{Planner} is an instance of an Agent specialized to
decompose a user instruction into several simpler tasks. It
provides a set of tasks and their dependencies in a structured format.
\end{quote}

\begin{quote}
The \textbf{Task Queue} is a container for holding the tasks that are created by the Planner. As tasks are completed, the Task Queue is responsible for propagating relevant results between tasks. It is also responsible for releasing tasks for execution once all of their dependencies have been resolved.
\end{quote}

\begin{quote}
The \textbf{Verifier} is an instance of an Agent tasked with
independent verification of the agent workflow result against the
original user instruction and the overall objective. The output from
the Verifier Agent is a boolean value indicating if the workflow result satisfies the request. If the result is deemed unsatisfactory, the workflow enters a replanning phase to improve the result.
\end{quote}

\begin{figure}[ht]
\centering
\includegraphics[width=0.7\textwidth]{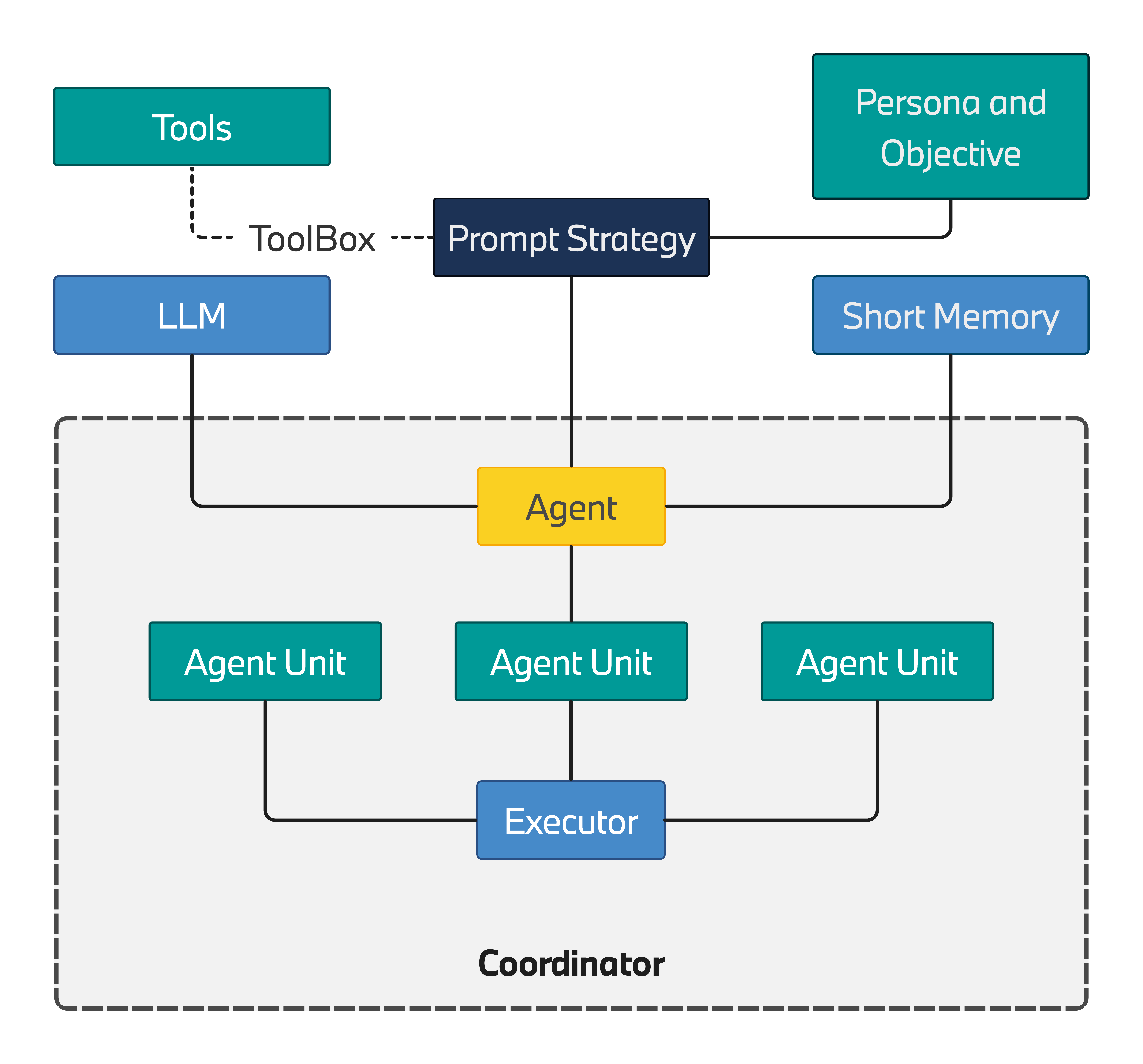}
\caption{Components contributing to initialization of an Agent. Dashed lines indicate optional
components.\label{figure-agent-creation}}
\end{figure}

The core of any Agent is an LLM, a prompt strategy, and a history of its past actions known as 
short memory, which will be described in detail in Section \ref{short-memory}. 
Figure \ref{figure-agent-creation} illustrates the components that make an Agent object and are 
involved in its execution. Although this diagram displays a single agent, this structure 
extends to multiple agents through the Agent Unit. 

Our prompt strategies are abstracted so that we can use both simple and iterative strategies without modification to an Agent. To demonstrate the standardization of prompts, in Subsections \ref{simple-prompt}--\ref{iterative-prompts} we will discuss a sample of non-iterative prompts, introduce tool usage and show the transition to the iterative ReAct prompt strategy \cite{react2023}, and its derivatives.

\subsection{Non-iterative Prompt Strategies}\label{simple-prompt}

Our most basic prompt strategy is a simple one-off LLM
call. In all of our strategies we include a template system
message with variables that can be modified. The default template includes
\textbf{persona} and \textbf{objective} variables 
and is stored in a plain text 
template file that is rendered when the prompting strategy is invoked.

Figure \ref{figure-non-iterative-prompt} 
displays message types used in the simple prompt strategy.
The ``system'' and ``user'' messages are inputs and ``assistant'' message represents 
the response from the LLM. We include a post-processing step that is responsible for parsing the raw model output based on the strategy and constructing a revised assistant message. This revised assistant message ensures that the iterative prompt strategies have consistent structure within short memory for the following iteration of LLM calls.

The basic strategy can be sufficient for components of the agent workflow which do not require iterative
processing with an LLM. Two examples of such Agents that can utilize basic strategies are the \textbf{Planner}
and \textbf{Verifier} Agents. These Agents can have specialized templates for their unique objectives and post-processing functions. In the following subsections we describe both
cases in more detail.

\subsubsection{Planner}\label{planner}

\begin{figure}[ht]
    \centering
    \subfigure[Basic]{
        \includegraphics[width=0.3\textwidth]{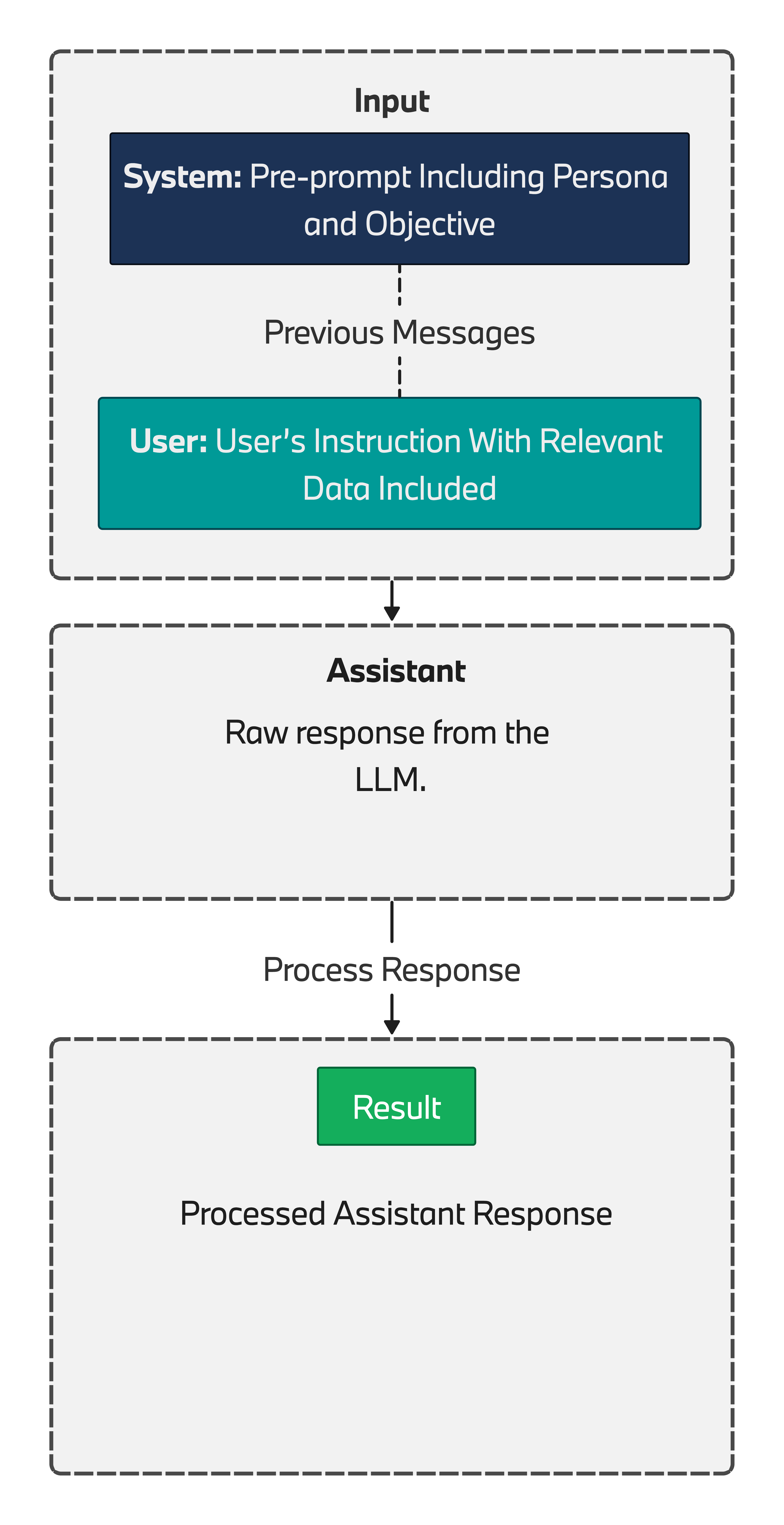}
        \label{figure-non-iterative-prompt}
    }
    \hfill
    \subfigure[Planner]{
        \includegraphics[width=0.3\textwidth]{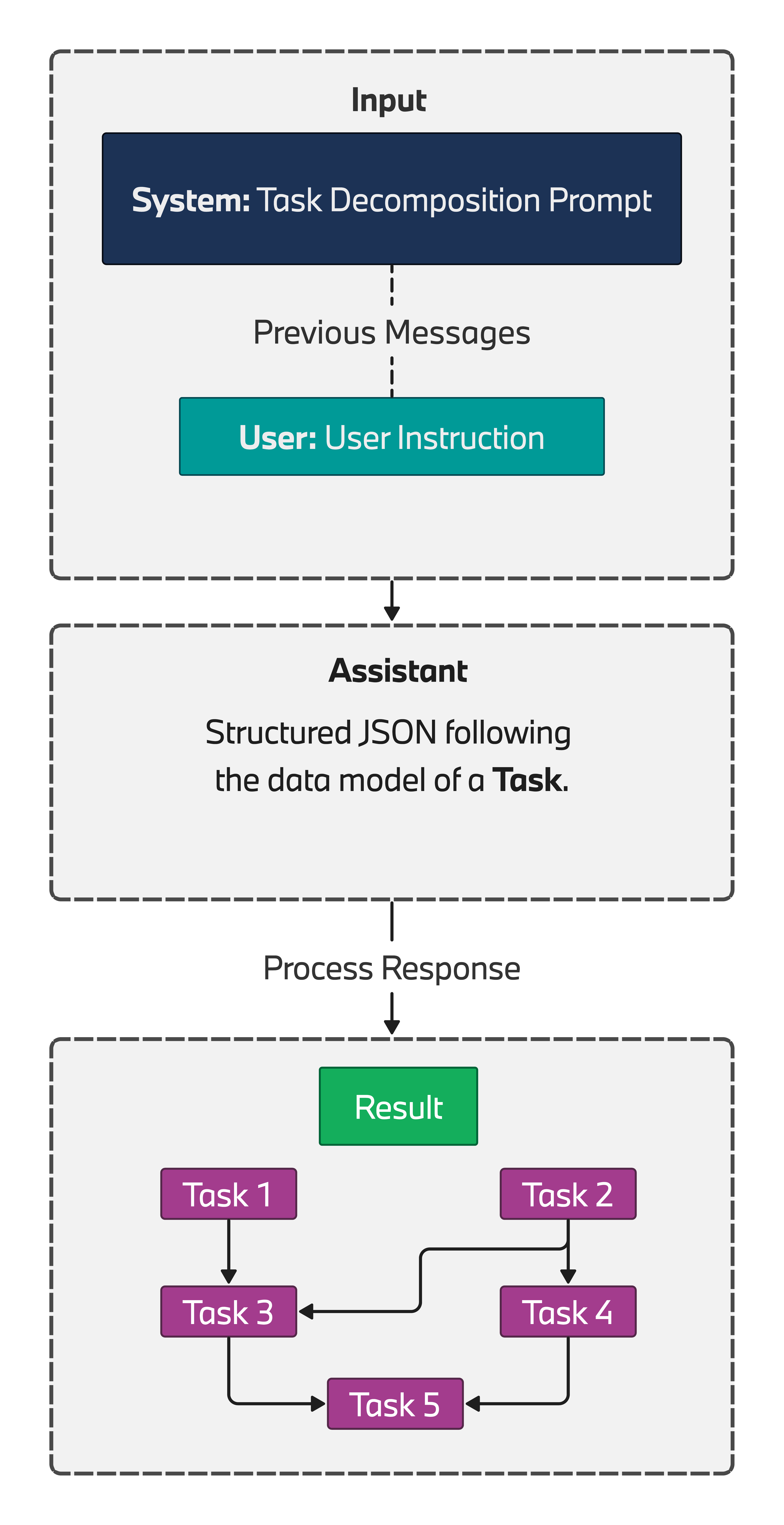}
        \label{figure-planning-prompt}
    }
    \hfill
    \subfigure[Verifier]{
        \includegraphics[width=0.3\textwidth]{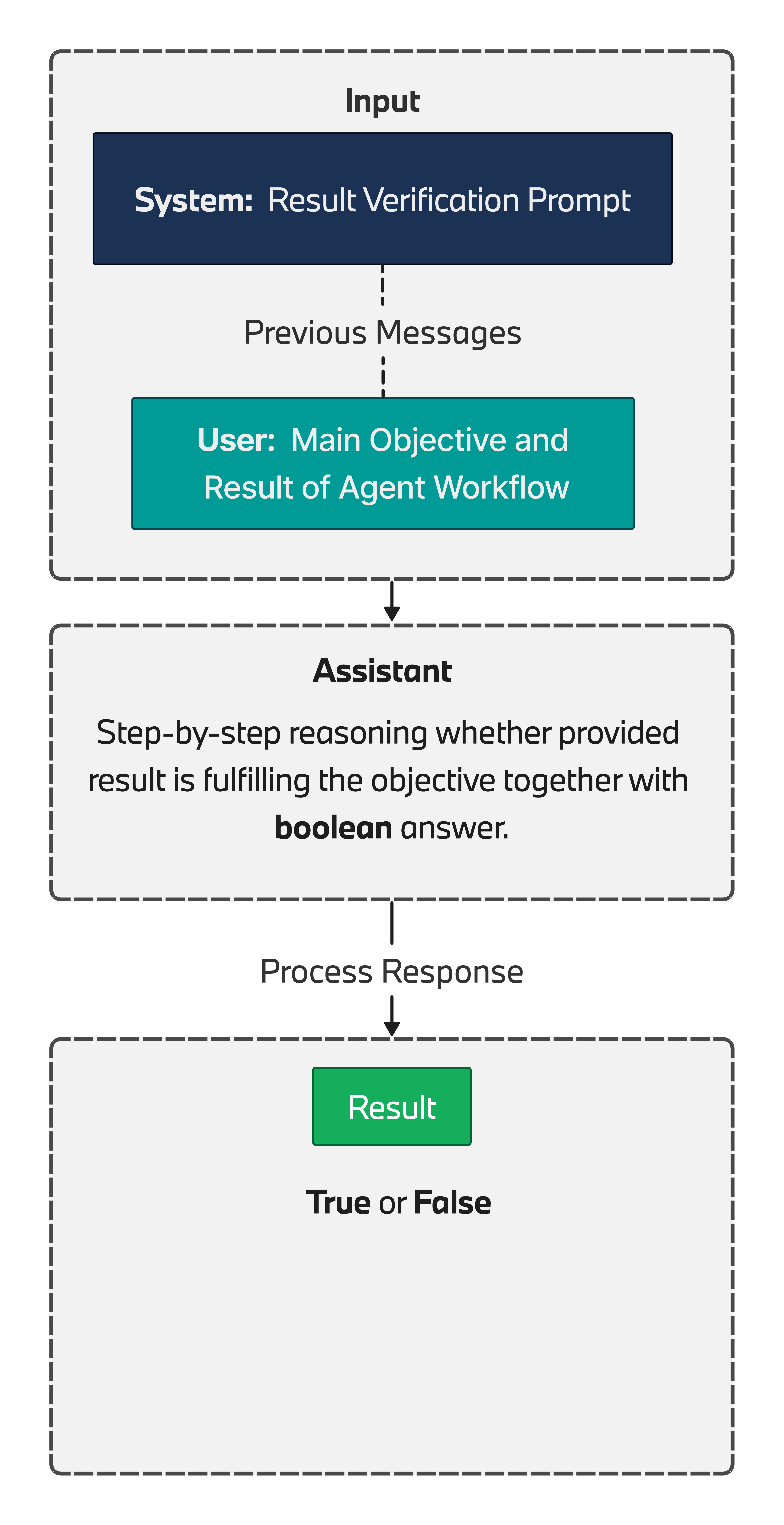}
        \label{figure-verification-agent}
    }
    \caption{Structure of \subref{figure-non-iterative-prompt} Basic
    non-iterative prompt, and its variations as applied in 
    \subref{figure-planning-prompt} Planner and 
    \subref{figure-verification-agent} Verifier agents.}
    \label{figure-all-non-iterative-prompts}
\end{figure}

An example that uses the non-iterative prompt strategy is simple task planning.
Task decomposition is a crucial element within a successful agent workflow.
The purpose of the Planner Agent is to take in the user's instruction,
decompose it into simple tasks, and identify dependencies between tasks in the form of a Directed Acyclic Graph (DAG). This simple, non-interactive
prompt strategy makes a single LLM call and does not further
refine its task list. The LLM is instructed to generate a response
in structured JSON format that will be parsed to extract all tasks and
populate the \textbf{Task Queue}. The data flow of this prompt strategy, from user instruction,
to the task DAG, to the final result is displayed in 
Figure \ref{figure-planning-prompt}.

\subsubsection{Verifier}\label{verifier}

Even with a planning stage, we
cannot guarantee that the agent workflow will be successful. A plan may not be completed properly or the created plan may not be sufficiently simple to be reliable. To provide a safeguard mechanism for autonomous task execution, we
include a Verifier Agent which is tasked with ensuring that the final
result sufficiently addresses the user's instruction. Figure \ref{figure-verification-agent} displays the structure 
of the Verifier Agent prompt when employing the simple prompt strategy. The Agent reasons through the result of the workflow and returns a true/false value. To prevent bias in the verification, it is done without knowledge
of the created plan or partial results completed during the execution.

\subsection{Tools}\label{tools}

It has been shown that LLMs can generate actions and automate tasks through the use of tools
\cite{nakano2022webgpt, ahn2022i, huang2022language}. These early demonstrations
have focused on web browsing or robotics but have been quickly 
generalized and are incorporated into the coherent ReAct prompt strategy
\cite{react2023}. Alternative approaches with extensive tool usage have been proposed 
in Gorilla \cite{patil2023gorilla} and others \cite{yuan2024easytool, song2023restgpt, tang2023toolalpaca}.

Tool usage is a crucial element of an agent workflow and our focus is on scalable and reliable usage. In this section, we discuss our approaches for ensuring a robust inclusion of tools within an agent's prompt strategy.

\subsubsection{Tool Abstraction}\label{tool-abstraction-and-hierarchy}

The presentation and execution of tools are implemented through a standardized interface that enables uniform agent interactions. In our work, we do not utilize the function calling feature of LLMs
provided by a limited number of APIs, such as OpenAI and Anthropic. 
Rather, we fully rely on the schema description for input and 
output as part of the system prompt.
We do not impose any simple common input schema for all the tools. Instead we define an exact list of parameters that a given tool requires as part 
of the prompt. Every tool in our implementation is equipped with a function
that provides its input and output schema. Each tool also includes a description 
that provides the LLM with the ability to identify the 
capabilities and purpose of a tool. 

\subsubsection{Toolbox Refiner}\label{toolbox-refiner}

While LLMs are capable of making an appropriate selection based on criteria, the accuracy can decrease as the number of options grows and limit the scalability of a system. Considering the range of tasks we wish to automate, the number of tools may be quite large. To maintain accuracy in the selection process, an agent should be presented a minimal, relevant subset of tools during its execution.

Our approach takes a philosophy of providing an agent with a wide
selection of tools, and at the same time minimizing the
amount of information provided in the system prompt for a given task.
This reduction of information is achieved by refining the set of tools
used by the agent in a given task by means of a \textbf{Toolbox Refiner}. The
task of the refiner is to start with a set of tools and, based on some specific
criteria, return a subset that will be presented to an agent for usage. Because the exact refining condition may differ based on the workflow, we utilized an interface for the refinement of tools. Our set of implementations
includes \textbf{Identity Refiner} which always returns the original set, \textbf{Hierarchical Refiner} which takes advantage of a 
hierarchy within tools through a tree search, and
\textbf{Semantic Refiner} which utilizes the semantic similarity between
tools' descriptions and task.

\subsection{Iterative Prompt Strategies}\label{iterative-prompts}

The non-iterative prompts are useful for tasks that do not require
any additional information from external sources to be completed. The main 
part of any agent workflow, however, will require interaction with the external
world through the use of tools. The result of the tool execution will be 
brought back as new information to the Agent before making another LLM call. In this section, we discuss ReAct and its derivatives to introduce iterative prompt strategies, tool usage, and bringing results back to the Agent.

\subsubsection{ReAct and PlanReAct}\label{react}

The ReAct prompt strategy \cite{react2023} provides an excellent example
of employing an LLM to reflect on a given task and interact
with a set of external actions. After the initial instruction, the ReAct strategy loops through Thought, Action and Observation stages, each serving a specific purpose. The \textbf{Thought} stage provides a place for the LLM
to reflect on the given task and necessary measures.
In the \textbf{Action} stage, the LLM defines the selected tool and appropriate input to this tool in JSON structure. The \textbf{Observation} stage does not occur inside of an LLM call, rather it is result of the tool execution with specified input and is returned to the LLM in the form of a User message. After this, the model goes
back to the Thought stage, equipped with more information as a result
of tool execution, to prepare for the next logical steps.

\begin{figure}[!h]
\centering
\includegraphics[width=0.7\textwidth]{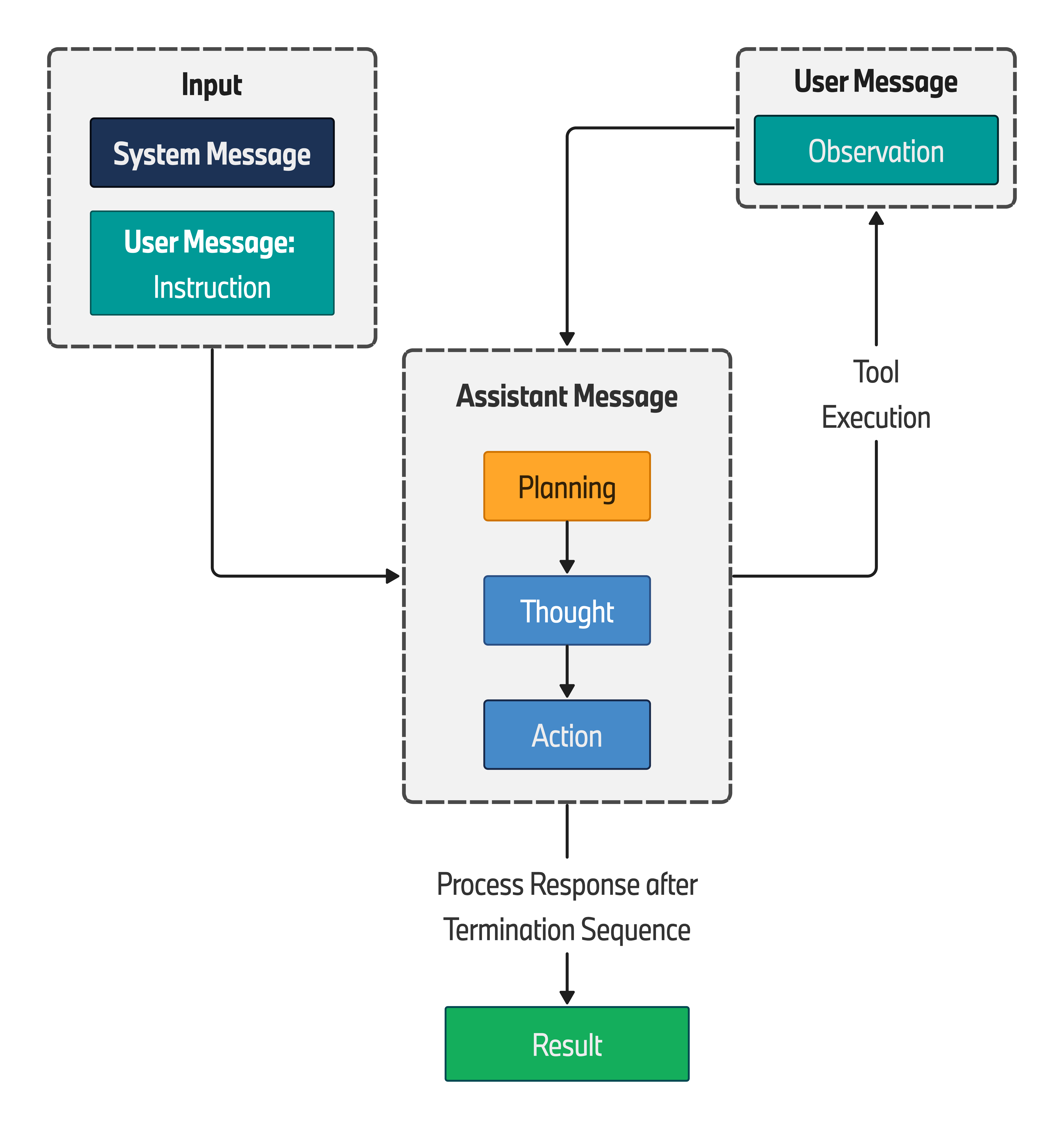}
\caption{Generic ReAct prompt strategy with Thought, Action and
Observation steps. We distinguish steps done as model responses
(Assistant Message) and as a user (including Observation as 
User Message). PlanReAct includes an additional Planning step marked in orange.\label{figure-react}}
\end{figure}

Figure \ref{figure-react} illustrates the steps employed in the iterative
structure of the ReAct prompt. Starting with the \textbf{System Message} and
an initial \textbf{Instruction} as the first \textbf{User Message}, we enter the loop of Thought and Action as the \textbf{Assistant Message} and
Observation as a \textbf{User Message}. We mark \textbf{User Message} and \textbf{Assistant Message} spaces to denote which
sections of the strategy are generated by an LLM and
which are placeholders for providing external information.

In the \textbf{System Message}, we describe how the model should respond if the
iterative sequence has ended and the final result is ready. In this case
the model is instructed to generate a defined termination sequence as part of the
response instead of new action. At every iteration we will
programmatically check if this termination sequence is present
and if found, break the execution loop. The final response is the content after the termination sequence.

We also implemented a specialized version of the ReAct prompt strategy that
explicitly includes a planning step as part of the iterative sequence. The
PlanReAct \cite{liu2023bolaa} strategy differs from the original ReAct by an 
additional step responsible for creating an explicit plan and replanning at 
every step. This prompting strategy is not meant to replace 
the Planner Agent, but rather works in tandem by providing 
an additional level of task decomposition in case the original task 
is still too complex.

\subsubsection{Programmable Prompt}\label{programmable-prompt}

\begin{figure}[!h]
\centering
\includegraphics[width=0.7\textwidth]{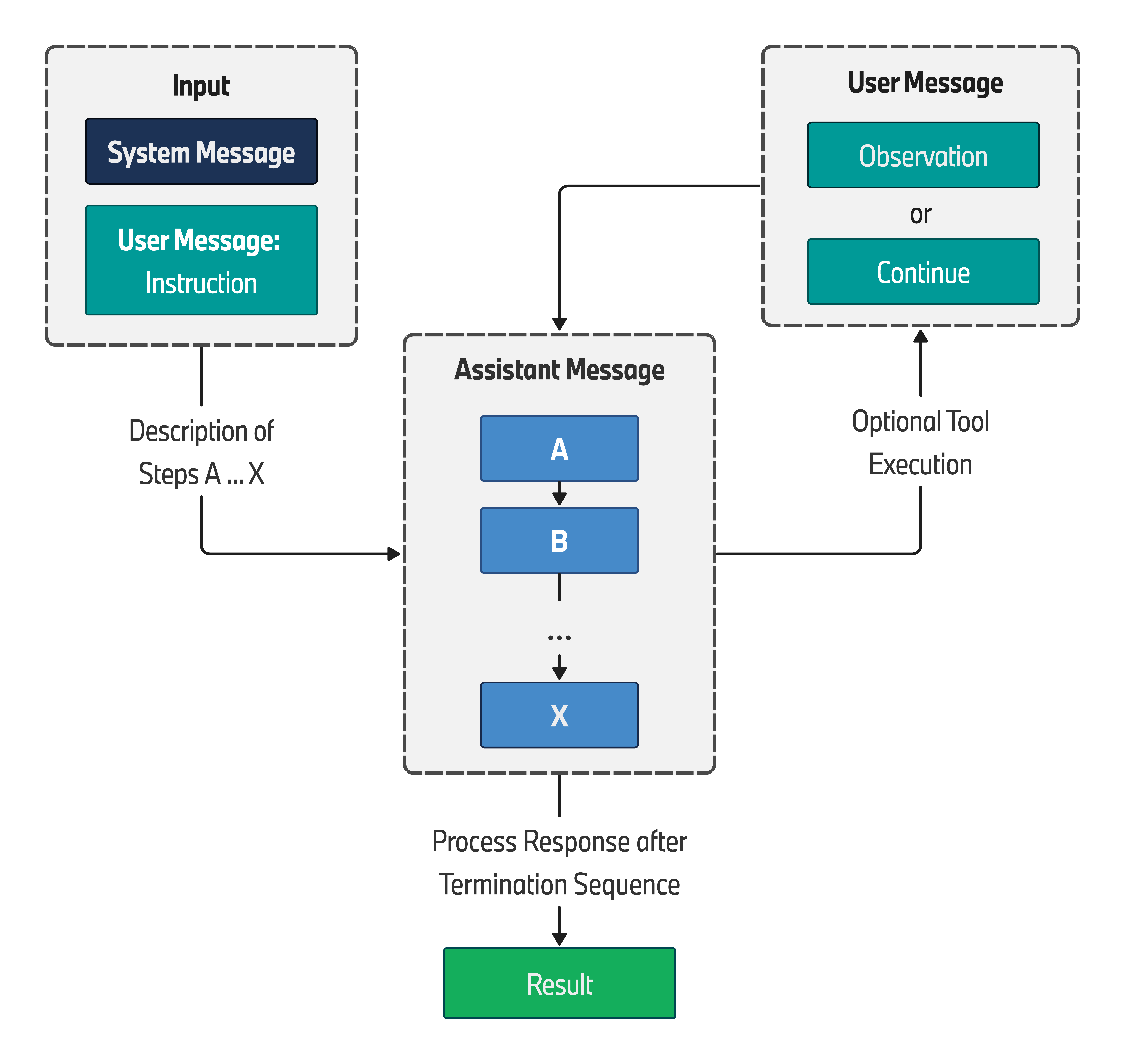}
\caption{Generic iterative prompt strategy with user defined steps
\texttt{A...X} that constitute the iteration sequence. The Assistant Message and
User Message show the combination of reasoning steps by the LLM and inclusion of external information.\label{figure-programmable-prompt}}
\end{figure}

We generalized the ReAct and PlanReAct strategies by introducing a configurable prompting strategy known as a Programmable Prompt.
The configuration will include an iterative sequence consisting of
predefined steps \texttt{A...X} where each letter represents a stage that
the LLM should take e.g.~\textbf{Thought}, \textbf{Action},
\textbf{Observation} for ReAct or \textbf{Plan}, \textbf{Thought},
\textbf{Action}, \textbf{Observation} for PlanReAct. This generalized 
prompting strategy is displayed on the diagram in Figure 
\ref{figure-programmable-prompt}.

The intention of this Programmable Prompt strategy, through the use of a 
configurable sequence, is to enable the employment of different decision making strategies. These strategies may mimic human decision making strategies such as Observe, Orient,
Decide, Act (OODA) \footnote{\url{https://en.wikipedia.org/wiki/OODA\_loop}},
Plan, Do, Check, Act (PDCA) \footnote{\url{https://en.wikipedia.org/wiki/PDCA}}, or similar derivatives and 
analogues.

\subsection{Data Propagation and Memory Levels}\label{data-propagation-and-memory-levels}

In order to have a full picture of how the initial instruction is solved,
we need to consider how the instruction is decomposed into simpler tasks
and how information propagates through the workflow. In the following
subsections we consider an example plan and trace the data needed to 
successfully complete all tasks. We will introduce types of task
dependencies and the methodology to  handle them. We also discuss how we employ memory concepts
\cite{packer2024memgpt, maharana2024evaluating, gao2024memory, 
wang2023augmenting} and distinguish the role and scope of short and 
episodic memory.

\begin{figure}[!h]
\centering
\includegraphics[width=1.0\textwidth]{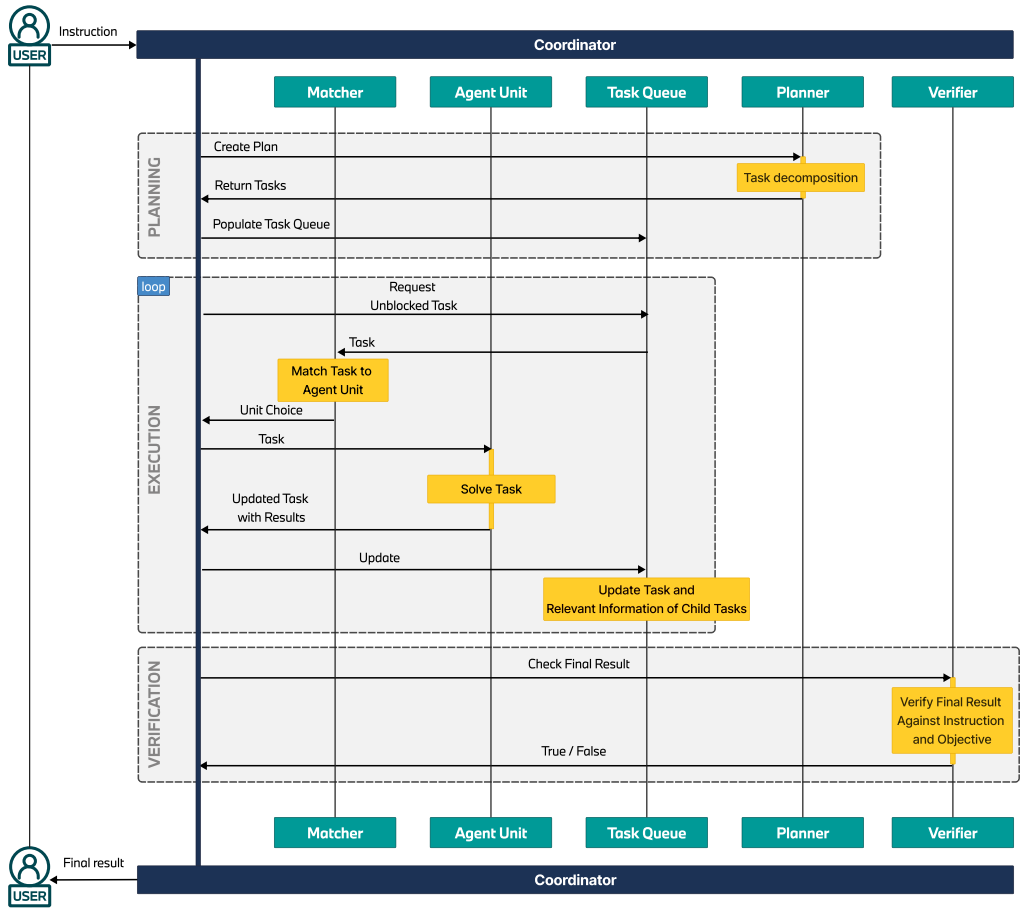}
\caption{Sequence diagram of operations in a generic agent workflow. We
display major components as vertical threads and type of operations
between them as an horizontal arrow. Three distinct blocks corresponding
to Planning (top), Execution (middle), and Verification (bottom) are
displayed on the diagram. For simplicity, we do not depict the replanning and re-execution that would occur if the Verifier returns false. \label{figure-sequence-diagram}}
\end{figure}

The sequence diagram in Figure \ref{figure-sequence-diagram} illustrates 
the order of operations in the Plan-Execute-Verify model introduced 
before in Section \ref{agent-workflow}. We display major elements of the 
workflow starting with the \textbf{Instruction} passed from the user to the 
Coordinator which orchestrates the entire workflow and provides the final 
result back to the user.

\subsubsection{Task Dependencies}\label{task-dependencies}

The Planner Agent's goal is to decompose the initial instruction into a
set of simpler tasks that are held in the Task Queue. In
our implementation, the Task object carries information about its
dependencies and results of those dependencies as they are completed. This approach is similar to the Graph of Thoughts
method by Besta et al. \cite{Besta_2024}.

\begin{figure}[!ht]
    \centering
    \subfigure[Direct dependency]{
        \includegraphics[width=0.7\textwidth]{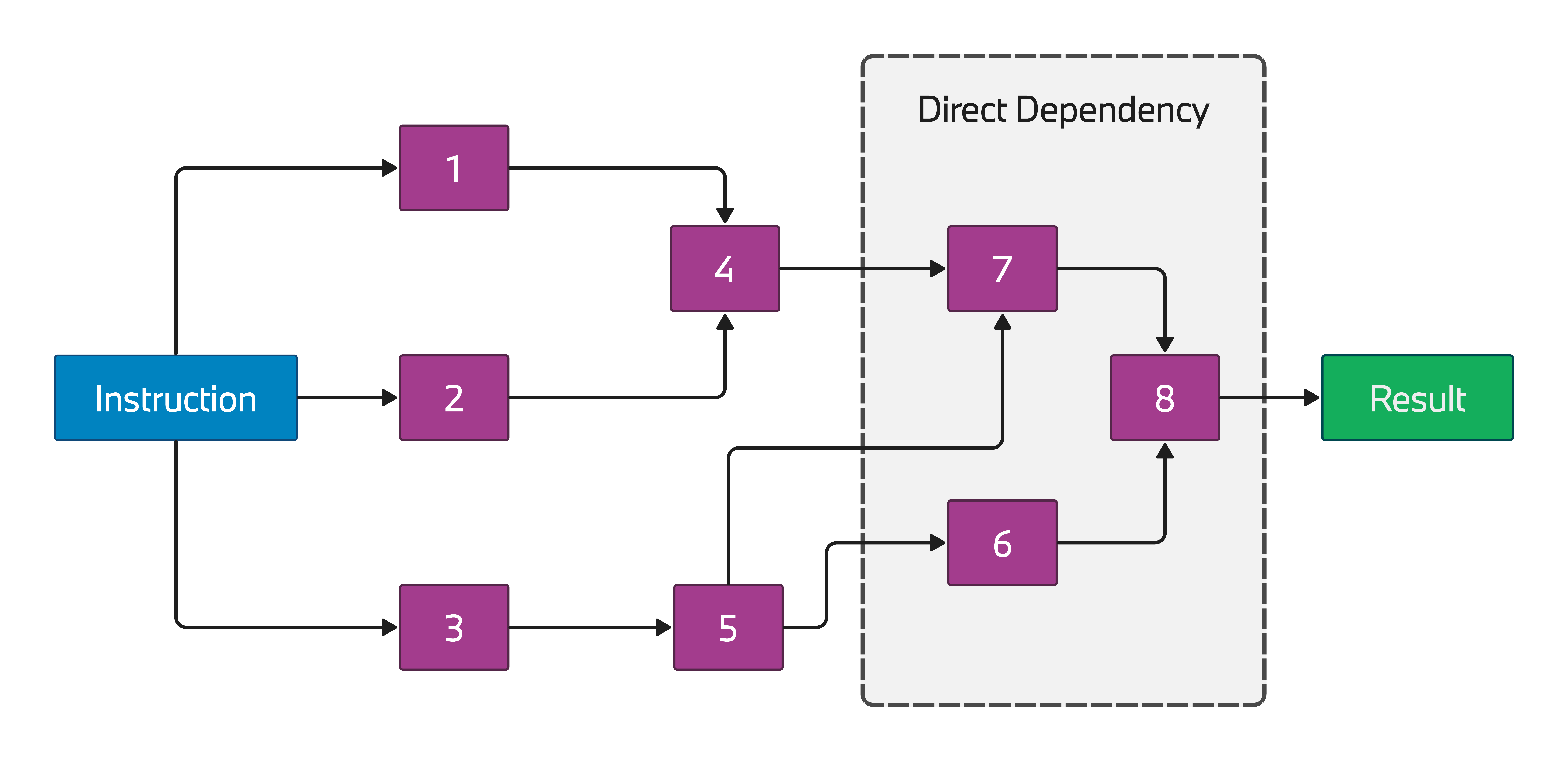}
        \label{figure-direct-dependencies}
    }
    \hfill
    \subfigure[Indirect dependency]{
        \includegraphics[width=0.7\textwidth]{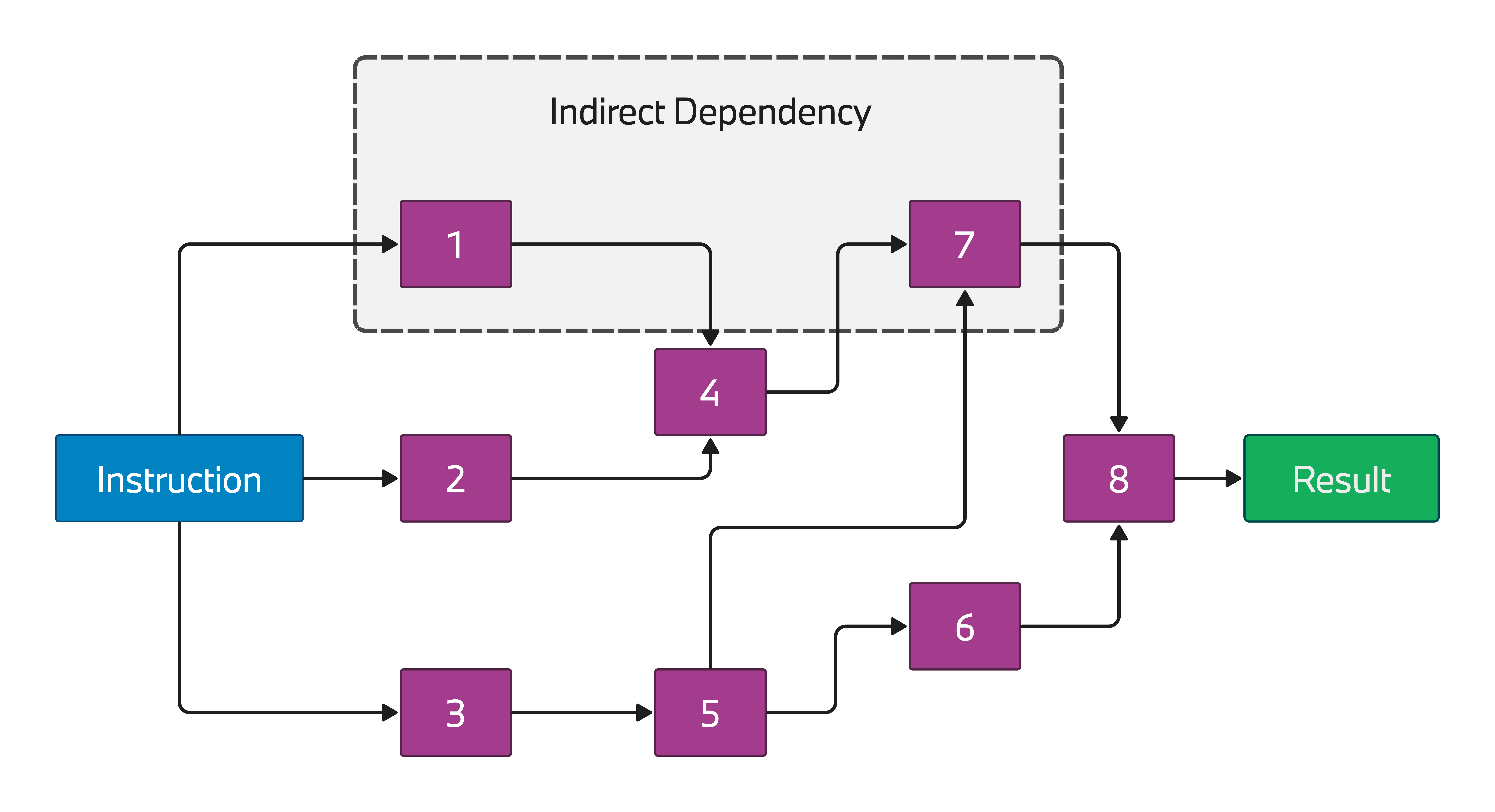}
        \label{figure-indirect-dependency}
    }
    
    \caption{Diagram showing a plan displayed as a DAG. 
        \subref{figure-direct-dependencies} Rectangle marks an example task 
        \textbf{8} with direct dependencies as \textbf{6} and \textbf{7}. 
        \subref{figure-indirect-dependency} Rectangle marks an example task 
        \textbf{7} and task \textbf{1} which is not a direct dependency but 
        is in the path of the execution leading to task \textbf{7}. 
    }
\end{figure}

In Figure \ref{figure-direct-dependencies} an 
example plan is displayed, starting with a user's \textbf{Instruction} 
leading to the \textbf{Result}. The initial instruction is decomposed into a set of tasks 
depicted as a DAG. Task dependencies are generated by the Planner during 
task decomposition. Direct dependencies are illustrated for an example task
\textbf{8}, which depends on tasks \textbf{6} and \textbf{7}. Results of tasks \textbf{6} and \textbf{7} are made available to an agent when solving task \textbf{8}.

However, we may define an indirect dependency as a result of any
preceding tasks in the plan that are not directly connected in the graph
to the current task. This indirect dependency is displayed on the
diagram in Figure \ref{figure-indirect-dependency} for the task 
\textbf{7} as a result of the task \textbf{1}. 
Task \textbf{1} has been chosen as random one from
the current plan that is not directly connected and precedes 
the task \textbf{7}. We need a special mechanism of
including the results of indirect dependencies for a current task. This
may be accomplished by semantic similarity of the previous results
(and/or task description) to the description of the task at hand. In
Subsection \ref{episodic-memory}, we introduce the concept of an \textbf{Episodic
Memory} that provides a solution for carrying over indirect
dependencies through the current plan.

The execution of each task is an independent action and no information,
other than the results of direct dependencies, are provided to an agent. In our implementation all tasks that 
are ready to be executed,
(all of their dependencies have been completed) are completed 
asynchronously. 

\subsubsection{Short Memory}\label{short-memory}

Depending on the employed prompting strategy, we may invoke a single call to an LLM (non-iterative prompt) or several calls (iterative prompts). To manage iterative prompts, we introduce a history of User (U) and Assistant (A) exchanges as
\textbf{Short Memory} (SM) where each generated message is stored in the
order of creation. The SM operates within the confines of a single task, remains isolated to each agent, and is purged upon task completion. This means that each agent's SM only contains messages that were generated for that agent. Optionally, we can enforce a maximum memory length by dropping oldest messages after the
capacity of the memory is exceeded.

\begin{figure}
\centering
\includegraphics[width=0.75\textwidth]{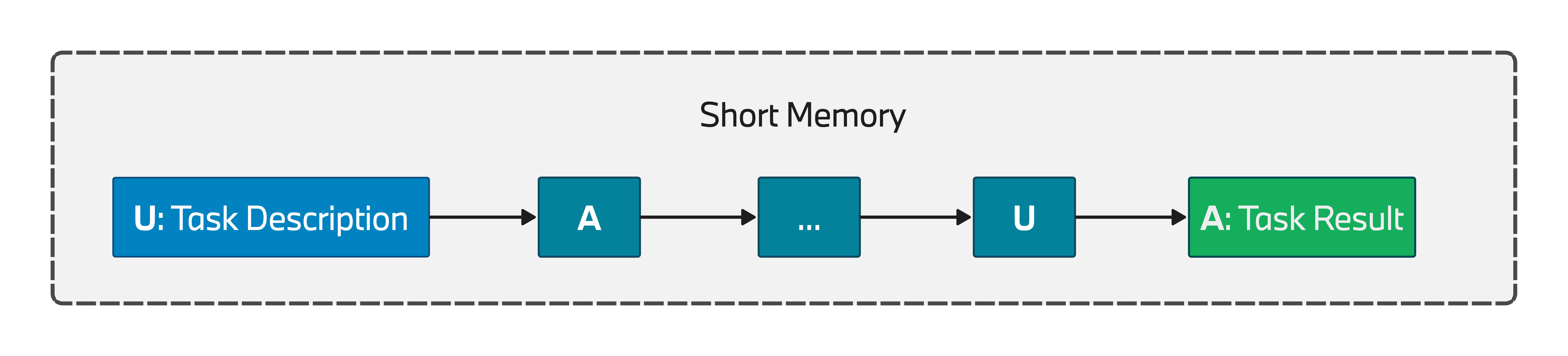}
\caption{Sequence of messages from the task description to the task result
including all intermediate messages. Rectangle marks the elements
contributing to the \textbf{Short Memory} of an agent. Messages are
marked in order of user inputs \texttt{U} and LLM responses \texttt{A}
(assistant).
\label{figure-short-memory}}
\end{figure}

Figure \ref{figure-short-memory} shows the sequence 
of messages in the order they are created. 
The square with ellipses marks the repeated pattern \textbf{U-\textgreater{}A} 
denoting \textbf{User (U)} and \textbf{Assistant (A)} type messages. 
The SM scope is marked with a grey rectangle.

\subsubsection{Episodic Memory}\label{episodic-memory}

In our implementation, \textbf{Episodic Memory} (EM) is a container that keeps
records of completed tasks across applications. This
container is a vector database to enable semantic retrieval.
Interactions with EM are done at the level of the Executor,
which has access to a task object for the entirety of its execution
cycle. Upon completion of a task, the Executor will package an \textit{Episode} consisting of the
task's description, result, and dependencies, in addition to
description and result embedding vectors,
and save it to EM. When handling subsequent tasks, the Executor can
semantically search for relevant Episodes from EM and provide them to an agent during task execution.

Episodic Memory attributes can be configured in several ways to
support the needs of the current workflow. For example, EM can be scoped to only retrieve Episodes from previously
completed projects or only from indirect dependencies of the
current project. Similarly, one can decide to retrieve only the
successfully completed episodes as well as other user-defined filters.

\begin{figure}
\centering
\includegraphics[width=0.75\textwidth]{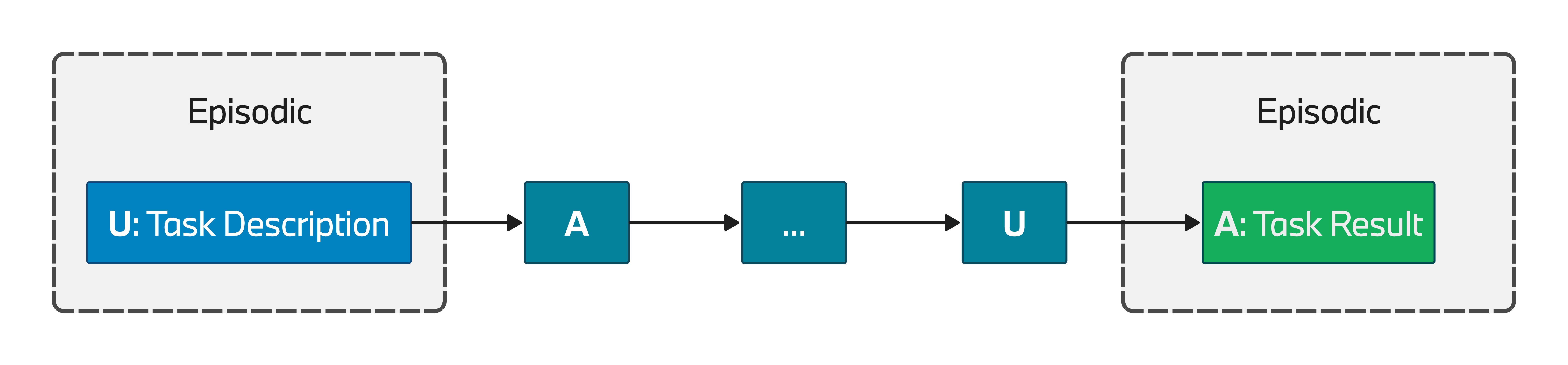}
\caption{Sequence of messages from the task description to the task result
including all intermediate messages. Rectangles mark the elements
contributing to the \textbf{Episodic Memory} of an agent. Messages are
marked in order of user inputs \texttt{U} and LLM responses \texttt{A}
(assistant).
\label{figure-episodic-memory}}
\end{figure}

In a similar fashion to the SM diagram, we visualize the EM content of a single task on the diagram in Figure \ref{figure-episodic-memory}. The scope of
EM, however, does not end with completion of the task.

Episodic Memory brings two important points:

\begin{enumerate}
\def\labelenumi{\arabic{enumi}.}
\item
  \textbf{Indirect Dependency Results} - In previous sections we have
  noted the need for bringing relevant results of previously completed
  tasks from the current plan. EM provides a solution to
  bring semantically relevant results from all previously completed
  tasks.
\item
  \textbf{Experiential Learning} - We can expand searches to all previously completed tasks, bringing results that are semantically
  relevant from the lifespan of all applications. This can enable a shortened path 
  to the final result, reducing execution time \cite{qian2024experiential}.
\end{enumerate}

\section{Multi-Agent Workflow}\label{multi-agent-workflow}

The need for reliability motivates a generalization of workflows from a single agent to
multiple agents. LLMs perform specialized tasks more reliably if they are
instructed to imitate certain personalities with specialized expertise.
The specification of expertise in the system prompt of an LLM removes any ambiguity when interpreting a given task. This
observation holds for simple human driven interactions and directly
translates into agent-based applications. Specifying the personality
of an agent, with focus on a narrow domain, leads to more reliable
performance that align with human preferences.

After defining a set of agents tailored for
solving a particular task, we may consider several strategies for
how a workflow may be orchestrated. The choices will indirectly mimic
human work organization for complex problems.

\subsection{Agent Unit and Agent Matching}\label{unit-execution-and-agent-matching}

In the general case of multi-agent workflows we face three issues, first the selection
of the Agent Unit for a given task, second the selection
of the agent to start the work on a given task, and third the dynamic selection of
the agent that will continue with next iteration. The crucial component for solving these issues has been briefly mentioned
in Section \ref{agent-workflow} as a matching mechanism. The \textbf{Matcher} is a generalized component that performs selection using a criteria
dedicated to the workflow e.g.~iterative matching, semantic matching, or mention matching
(as described in details below using @ notation). After the Planner creates a set of tasks, we use a matching function to determine which Agent Unit should solve a given task. Because each Agent Unit consists of one or more agents, we also need a mechanism for selecting an agent for each iteration of the task execution. Before executing an iteration of the 
agent prompt strategy, we again invoke a matching function to select the agent that 
will be used for this iteration. 

The introduction of an \textbf{Agent Unit} and a \textbf{Matcher} allows us to have a
flexible method for considering single and multi-agent workflows using
the same implementation. Although single agent workflows could be accomplished without these components, they are crucial for multi-agent
applications. All of the described multi-agent strategies become
a particular combination of prompt strategy, Agent Unit, and matching
function. For example, a two agent actor/critic workflow may be realized by
introducing a unit of two agents i.e. \textbf{actor} and \textbf{critic}, and
an iterative matching function that cycles through the Agent Unit. 

In Subsection \ref{conversational-prompt-strategy} we will introduce dynamic selection of the next
agent by explicitly mentioning its name in the current agent response
with \texttt{@AgentName} notation. This notation gives an indication
which agent should be selected next and a mention matching
function will provide the correct agent from the available Agent Unit.

\subsection{Workflows with Multiple Agents}\label{workflows-with-multiple-agents}

In this section, we consider five different strategies for multi-agent
workflows that are enabled through the components \textbf{Agent Unit} and \textbf{Matcher} and are applicable to the \textbf{Plan-Execute-Verify} fashion of
solving a complex task, 1) Independent, 2) Sequential, 3) Joint, 
4) Hierarchical, and 5) Broadcast. In each, we consider only the
\textbf{Execute} part of the workflow and assume that the
Coordinator is solving a set of tasks organized as a DAG. All strategies are illustrated
with an example in Figures \ref{figure-independent-and-sequential}, 
\ref{figure-hierarchical-and-joint}, and \ref{figure-broadcast}. 

\subsubsection{Independent}\label{independent}

In this strategy, we consider tasks to be executable by a single
agent. The coordinator orchestrates the solution to a set of tasks organized in a DAG
using an Agent Unit composed of several agents, each with a
distinct purpose. If the set of tasks is sufficiently simple and matches the narrow expertise of a single agent from the unit, we may assume that this agent will be able to
solve the given task.

Within an Agent Unit, a task will be assigned to a single agent using some matching logic which was detailed in the previous section. Each task will be executed by a single agent in isolation
of other agents and no inter-agent communication will take place during
the task execution. However, the agents may still have access to the results of other completed
tasks based on the direct and indirect results
propagation, by either explicit passage of results as direct
dependency or usage of EM.

Figure \ref{figure-independent-and-sequential} 
illustrates the \textbf{Independent} model on
panel (1) with an example of a graph with seven tasks and three agents.
Dependencies of tasks are not displayed as  tasks will only be executed when all dependencies are solved.

\subsubsection{Sequential}\label{sequential}

In many circumstances, we may consider a workflow that requires 
consecutive steps where several distinct agents would perform better than a
single one. An example of this may be student/teacher 
workflow where one agent performs the task and the second provides constructive feedback and suggestions. We may generalize
this example to include more than two agents and a workflow consisting of
more than two consecutive steps.

The \textbf{Sequential} strategy considers a single task 
which is executed by an Agent Unit consisting of two or more agents. This
execution will be performed in a sequence of consecutive steps where
each may be done by a different agent. The student/teacher example
would correspond to a unit of two agents with alternating
order. The Sequential strategy imposes a predefined sequence of agents rather than an autonomous decision on which agent to pass to next.

\begin{figure}[!ht]
\centering
\includegraphics[width=1.0\textwidth]{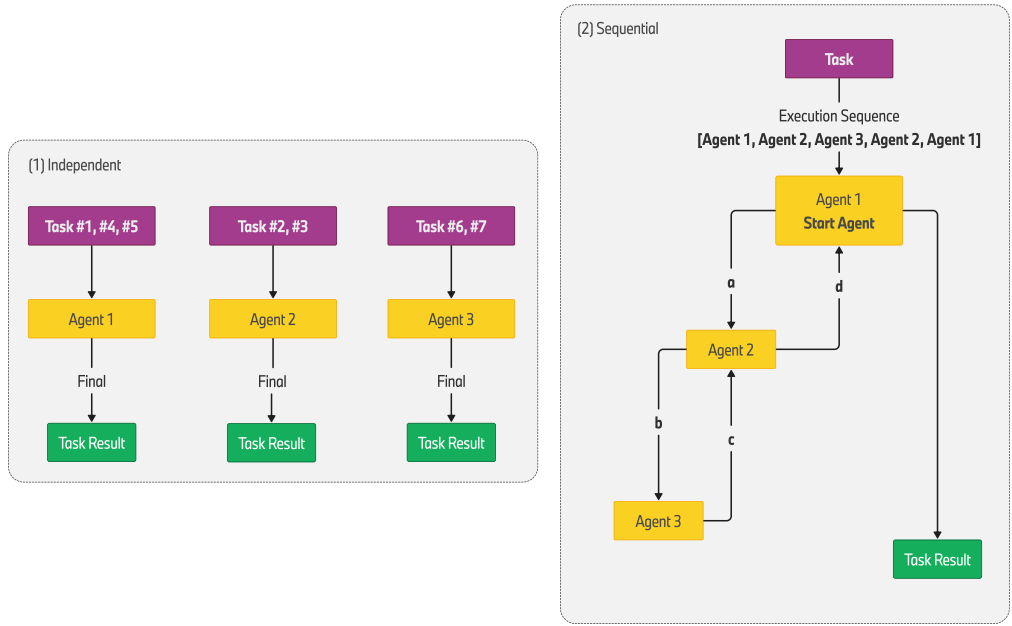}
\caption{Multi-agent strategies. (1) \textbf{Independent} with individual tasks (numbered from \#1 to \#7)
among three agents named Agent 1, Agent 2, and Agent 3. Each task is
considered to be completed by a single agent with no communication during
the task execution. Results of agent work are only shared through the
direct and indirect dependencies and results propagation. (2)
\textbf{Sequential} with a single task executed by a unit of agents consisting of Agent 1, Agent 2, and Agent 3 with a specific
sequence of execution. In the displayed example the sequence of execution is
{[}Agent 1, Agent 2, Agent 3, Agent 2, Agent 1{]}. The task execution starts with Agent 1 and after it
completes work, Agent 2 starts with the partial result of Agent 1.
\label{figure-independent-and-sequential}
}
\end{figure}

In Figure \ref{figure-independent-and-sequential} panel (2), we illustrate a Task X being solved
by the unit of three agents in the predefined order of {[}Agent 1, Agent 2, Agent 3, Agent 2, Agent 1{]}.
A single step is completed by a given agent by generating an LLM response. The following iteration is completed by the next
agent from the Agent Unit based on the sequence imposed by the matching strategy.

\subsubsection{Joint}\label{joint}

The \textbf{Joint} pattern resembles the collaboration between a group of peers. In this pattern, all agents have knowledge of all other agents and are allowed to trigger the final result of a workflow. The execution of a task begins with a predefined agent. However, in each iteration of the prompt strategy, an agent solves a component of the task and then decides whether to pass the execution or trigger the final result. If execution is passed, the current agent will select the following agent from all agents available (including itself). In this way every agent can communicate with every other agent without a prescribed order.

In Figure \ref{figure-hierarchical-and-joint}, 
we illustrate the \textbf{Joint} workflow. The example shows how Task X can be solved by an
autonomous team. The example task is started by Agent 1 which has been determined to be the most relevant agent to start working on this task. In the next iteration, Agent 1
will decide which agent should take over. This may include
self-selection. In our example, Agent 3 terminates and provides the final result.

\subsubsection{Hierarchical}\label{hierarchical}

The \textbf{Hierarchical} strategy mimics a manager/worker pattern where a particular agent is in charge. Only this agent may distribute the work but it is also allowed to complete parts of it by itself. We define a lead agent for a given task that initially receives the task and is aware of all other agents in the unit. The lead agent is responsible for solving the task with the aid of other agents in the Agent Unit by passing the execution to another agent for one or more steps of the original
task. Non-lead agents are only informed about the existence
of the lead agent and are not allowed to trigger the completion of a task. Therefore, any non-lead agent in the unit may only report the results of their work back to the lead agent. 

\begin{figure}[!ht]
\centering
\includegraphics[width=1.0\textwidth]{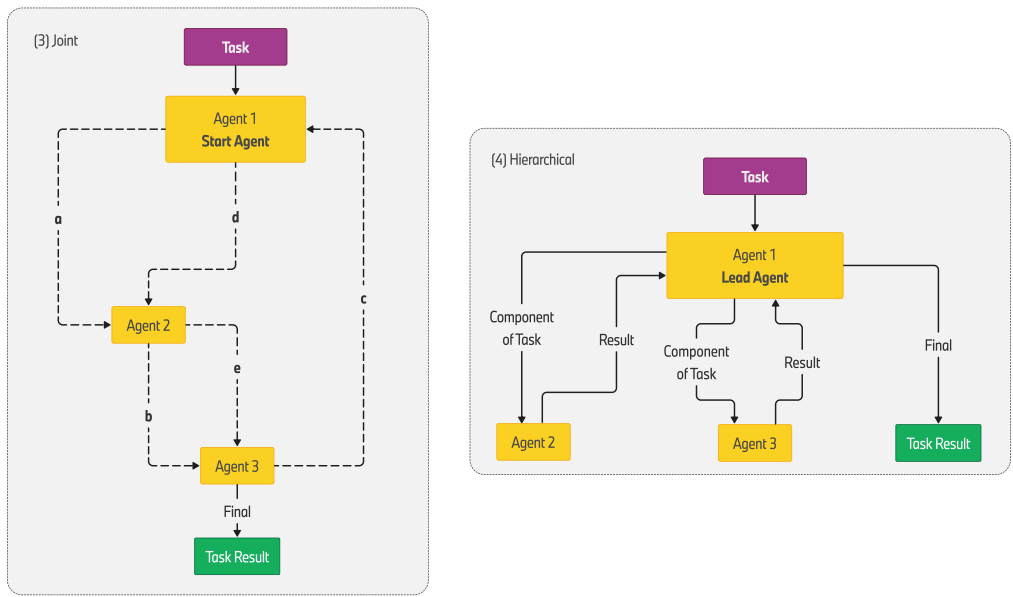}
\caption{Multi-agent strategies. 
(3) \textbf{Joint} with Task X being executed by an Agent Unit consisting of three agents
named Agent 1, Agent 2, and Agent 3. Agent 1 is selected as the starting
agent. All agents in the unit are allowed to communicate with each other and each agent is allowed to end the workflow when complete. In
both \textbf{Hierarchical} and \textbf{Joint} cases, the next agent is
selected dynamically and no fixed sequence of execution is imposed.
(4) \textbf{Hierarchical} with
Task X being executed by an Agent Unit consisting of three agents named Agent 1,
Agent 2, and Agent 3. Agent 1 is the lead agent in this unit and is
aware of the other two agents. Agent 1 is also the only one that is allowed
to terminate the workflow and provide the final result. Agent 2 and Agent 3
can communicate only with Agent 1 and are not allowed to terminate. 
\label{figure-hierarchical-and-joint}}
\end{figure}

In Figure \ref{figure-hierarchical-and-joint} we display an example workflow where a unit of three agents is working on Task X. Agent 1 is selected as the lead agent and passes the execution of task components to Agents 2 and 3. In this process, the target agents are autonomously decided by Agent 1. The results of each task step are passed back to Agent 1, which decides when to trigger the completion of the full task.

\subsubsection{Broadcast}\label{broadcast}

\begin{figure}[!ht]
\centering
\includegraphics[width=0.5\textwidth]{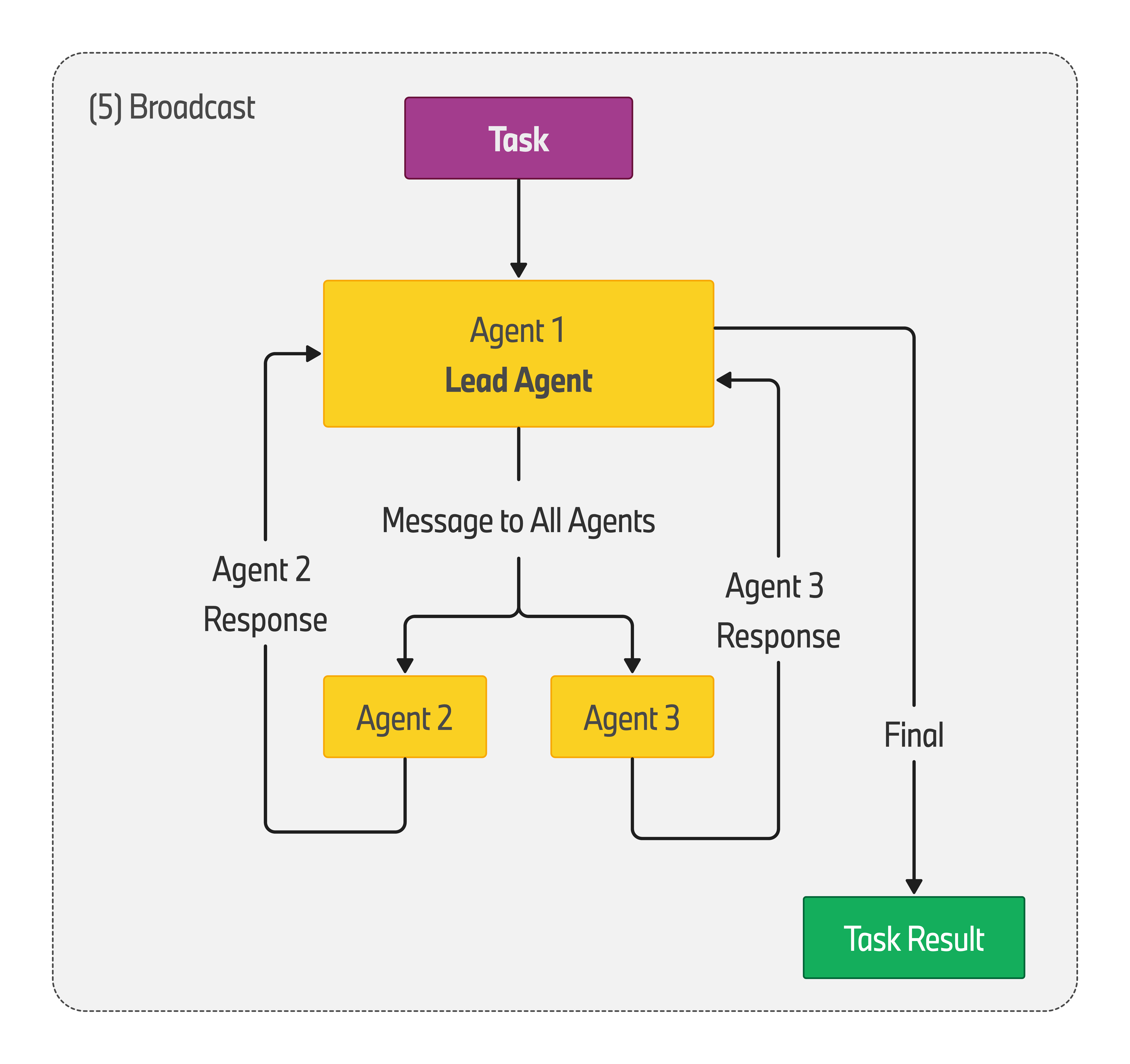}
\caption{Multi-agent strategies. (5) 
The \textbf{Broadcast} method is
similar with the \textbf{Hierarchical} approach with the difference that
the lead agent broadcasts the same message to all other agents in the
unit and those respond independently to the lead agent. In this example,
Agent 1 is aware of all other agents and solicits the response from all
of them in the same fashion. 
\label{figure-broadcast}}
\end{figure}

The \textbf{Broadcast} method is a derivative of the
\textbf{Hierarchical} approach. The main difference is
that the lead agent carries individual conversations with all other agents by broadcasting the same message to the whole Agent Unit. Non-lead agents are not aware of the existence of others and operate independently, providing individual responses back to the lead agent. The lead agent waits for all responses to proceed with the next
iteration. At the following iterations, the lead agent may broadcast
another message or terminate with a final result. As illustrated in Figure \ref{figure-broadcast}, only the lead agent has full information about all responses and can see the full group conversation. All non-lead
agents carry independent conversations.

\subsection{Conversational Prompt
Strategy}\label{conversational-prompt-strategy}

\begin{figure}
\centering
\includegraphics[width=0.7\textwidth]{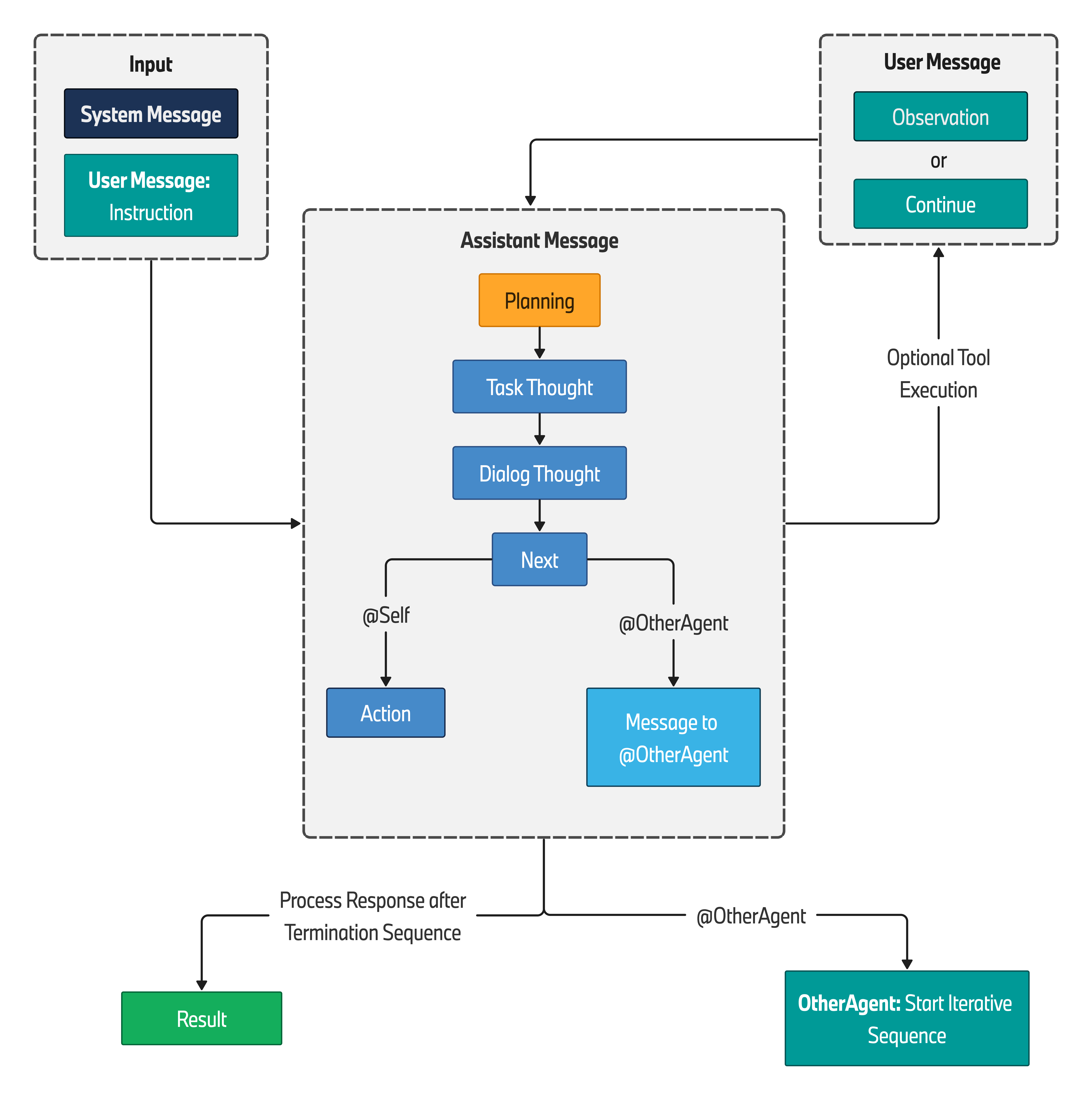}
\caption{Iterative sequence of ConvPlanReAct prompting strategy. The
ReAct and PlanReAct strategies are generalized to include additional
steps related to dialog with another agents. The \textbf{Thought}
stage is split into \textbf{Task Thought} and \textbf{Dialog Thought} stages. The
\textbf{Next} stage is introduced for indicating which agent should
continue with the execution of the current task using
\texttt{@AgentName} notation. \label{figure-convplanreact}}
\end{figure}

LLMs, and in particular their chat optimized versions, are
inherently designed to follow pairwise interactions. This pairwise
interaction has been generalized to enable automated interaction with the
external world using prompt strategies like ReAct \cite{react2023} that includes Action/Observation pairs. While ReAct and PlanReAct enable interaction with external functions, they do not enable any dialog at the Action and Planning steps such as the techniques proposed to
enable multi-agent dialog \cite{li2023camel, rasal2024llm, wang2024unleashing, 
fu2023improving, liu2024llm}. To address this issue, we employ a strategy that aims to generalize
PlanReAct prompting and enable a dynamic dialog between many agents, illustrated in Figure \ref{figure-convplanreact}. We
select an iterative method that builds on ReAct for a few reasons. First,
the ReAct loop provides a stable execution pattern that is reliably
followed by LLMs in consecutive iterations. Second, it
enables the natural form of Action/Observation and the ability to execute external
functions. Third, the iterative sequence is resilient to small
deviations such as missing new data in the Observation stage or using ``Continue''
statements instead of providing additional information. Fourth, generalization
of ReAct to PlanReAct makes it capable to work on more complex tasks with longer
execution sequences. To further extend capabilities to include dialog, we introduce the following changes:

\begin{itemize}

\item
  The \textbf{Thought} stage is split into two sub-stages separating the
  reflections related to Task and Dialog. We introduce the \textbf{Task
  Thought} stage that encourages the agent to reflect on the current task and the next
  step to solve it, and the \textbf{Dialog Thought} stage where the agent can
  reflect on how to best utilize other agents for solving the current
  problem. Information about other available agents is provided to
the agent as part of the system prompt, in similar syntax to the one
mentioned before for tool usage in Section \ref{tools}.
\item
  A \textbf{Planning} stage is included, and the next steps to solve the 
  current task are reviewed and re-planned in every iteration.
\item
  The \textbf{Next} stage is added to the iterative sequence to indicate which agent from the available Agent Unit should
  continue solving the current task. We instruct the LLM to use
  \texttt{@Self} if the current agent is best choice, or indicate the name
  of another agent using the notation \texttt{@OtherAgent}.
\item
  The \textbf{Action} stage is only included if the next agent is \texttt{@Self}.
\end{itemize}

\subsection{Human Feedback}

In multi-agent strategies we need to 
consider a method for including human feedback
to agents while task execution is in progress. We consider
two types of human feedback from the perspective of the autonomous
agents work:

\begin{itemize}

\item
  \textbf{Intentional} - The Agent Unit may include a Human Proxy
  agent that is responsible for passing a request from the agents to a
  human proxy while working on a task. From the agent's perspective this
  is no different than passing the execution to another agent in the
  Agent Unit. The current agent may decide to request
  additional information and pass the request to a Human Proxy by mentioning
  \texttt{@HumanProxy} in the generated response. However,
  from the programmatic view, this requires consideration of the human response time and a potential pause of the Agent Unit execution.
\item
  \textbf{Incidental} - While the explicit request for feedback is handled in the prompt strategy by including a HumanProxy
  agent, there may be occasions when a human monitoring the workflow decides that the workflow execution deviates from the desired outcome. There
  needs to be a different mechanism for providing this \emph{in situ}
  feedback to the executing agent as an unsolicited message from the
  HumanProxy agent. The prompt strategy does not have a natural place
  to bring such an interaction other than the \textbf{Observation}
  stage. At that stage we may bring additional information about the
  external world to the agent such as the result of
  the requested action. However, the prompt strategy is resilient to
  deviations from the assumed flow. For example, the strategy can include the phrase \emph{Continue} as the observation which leads to a continuation of the sequence's next stage. The incidental human
  feedback may be brought to the agent in the \emph{in situ} fashion as
  one of the observations and the additional information will be
  available at the next iteration of the current task execution.
\end{itemize}

\section{Example Applications}\label{example-applications}

To demonstrate the flexibility of our approach, we highlight three different example applications and how each can be achieved using the components described in previous
sections. We will start with a Question \& Answer workflow capable
of answering complex multipart questions. This workflow is based on
a single agent utilizing a semantic search tool. Second, we will describe
a document editing workflow with two agents in an actor/critic iterative
mode. Third, we will show how a software development workflow may be
organized utilizing three agents each using different tools to interact
with external systems.

In the diagrams used in the following sections, we use solid lines to 
show programmatically imposed flows and dashed lines to display optional flows that 
are decided by the agents.

\subsection{Question \& Answer - Retrieval Augmented
Generation}\label{question-answer---retrieval-augmented-generation}

Retrieval Augmented Generation (RAG) systems have proved to be a promising
solution to bring additional information to an LLM. RAG methods are particularly beneficial to industrial
applications as they provide a solution for internal knowledge without the need to fine-tune or retrain an LLM.
This ensures that the model may always access relevant
information with minimal maintenance for productive
deployment.

\begin{figure}[!ht]
\centering
\includegraphics[width=0.7\textwidth]{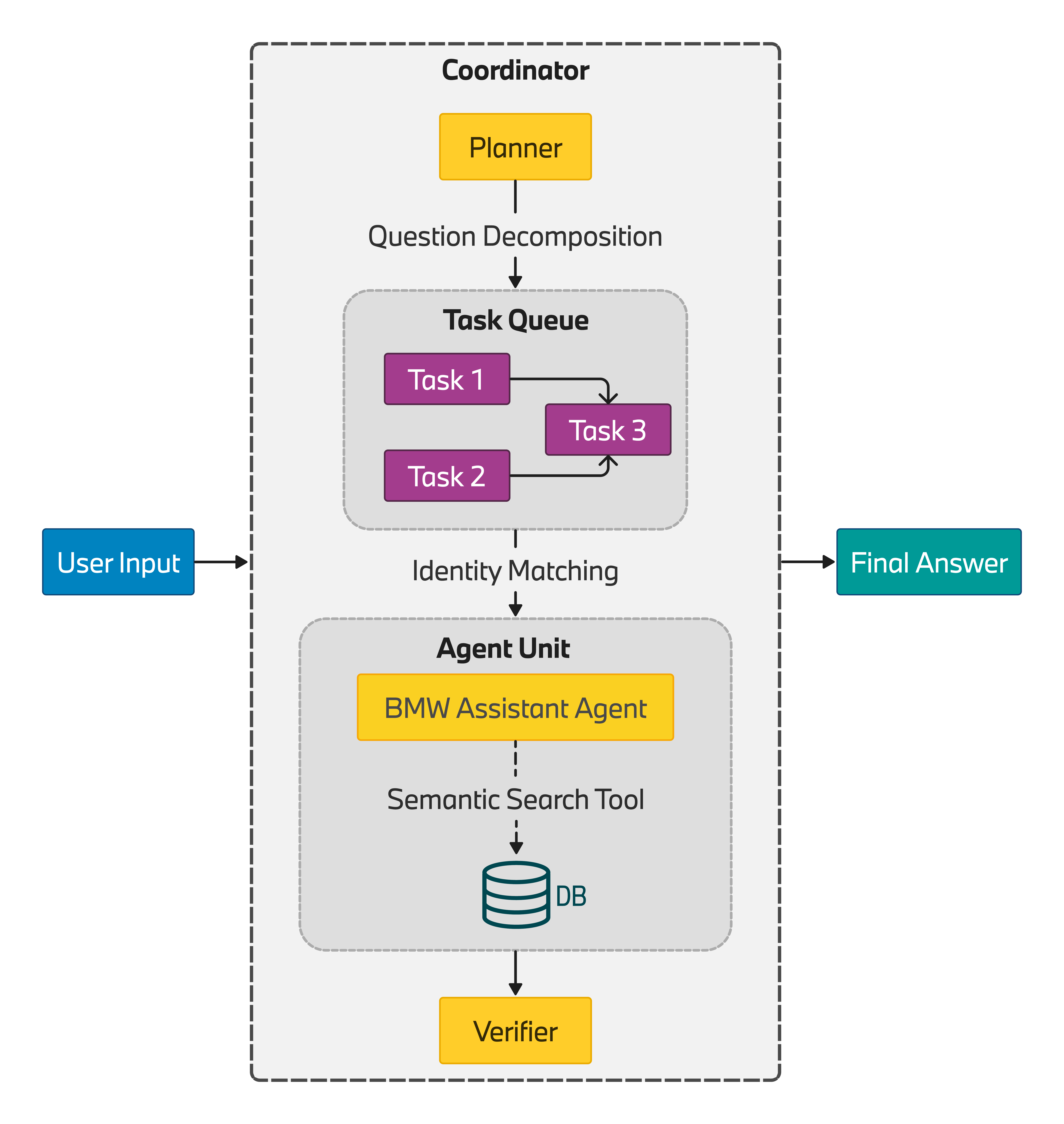}
\caption{Agent workflow for Retrieval Augmented Generation for a
Question \& Answer system. The workflow engages a single agent with access to a semantic search tool. This agent executes
all tasks from the plan that consists of simple questions originating
from the task decomposition of the original user question.
\label{figure-rag-example}}
\end{figure}

There have been many RAG algorithms proposed to improve on the naive RAG approach. In this example, we focus on an agent-based realization of
the RAG method for question-answering systems. The workflow displayed 
in Figure \ref{figure-rag-example} is
realized with a single agent, based on either the ReAct or the PlanReAct prompt
strategy. 

The workflow starts with a user question to the
\textbf{Coordinator} who passes it to the \textbf{Planner} for task decomposition, breaking the complex question into a set of
simpler tasks with dependency mapping. The decomposition of the initial user question ensures that complex
questions are first reasoned through and simplified into set of questions
that contribute to the final answer. Tasks are then
executed by a single agent denoted as \textbf{BMW
Assistant} who has access to a \textbf{Semantic Search} tool for identifying relevant information. However, this toolbox can be extended with tools for accessing other internal systems such as web sites or enterprise
knowledge management systems. The final answer of the agent is verified at the end of the workflow by the
\textbf{Verifier} agent to ensure the final response addresses the original user question.

\subsection{Document Editing -
Actor/Critic}\label{document-editing---actorcritic}

\begin{figure}[!ht]
\centering
\includegraphics[width=0.7\textwidth]{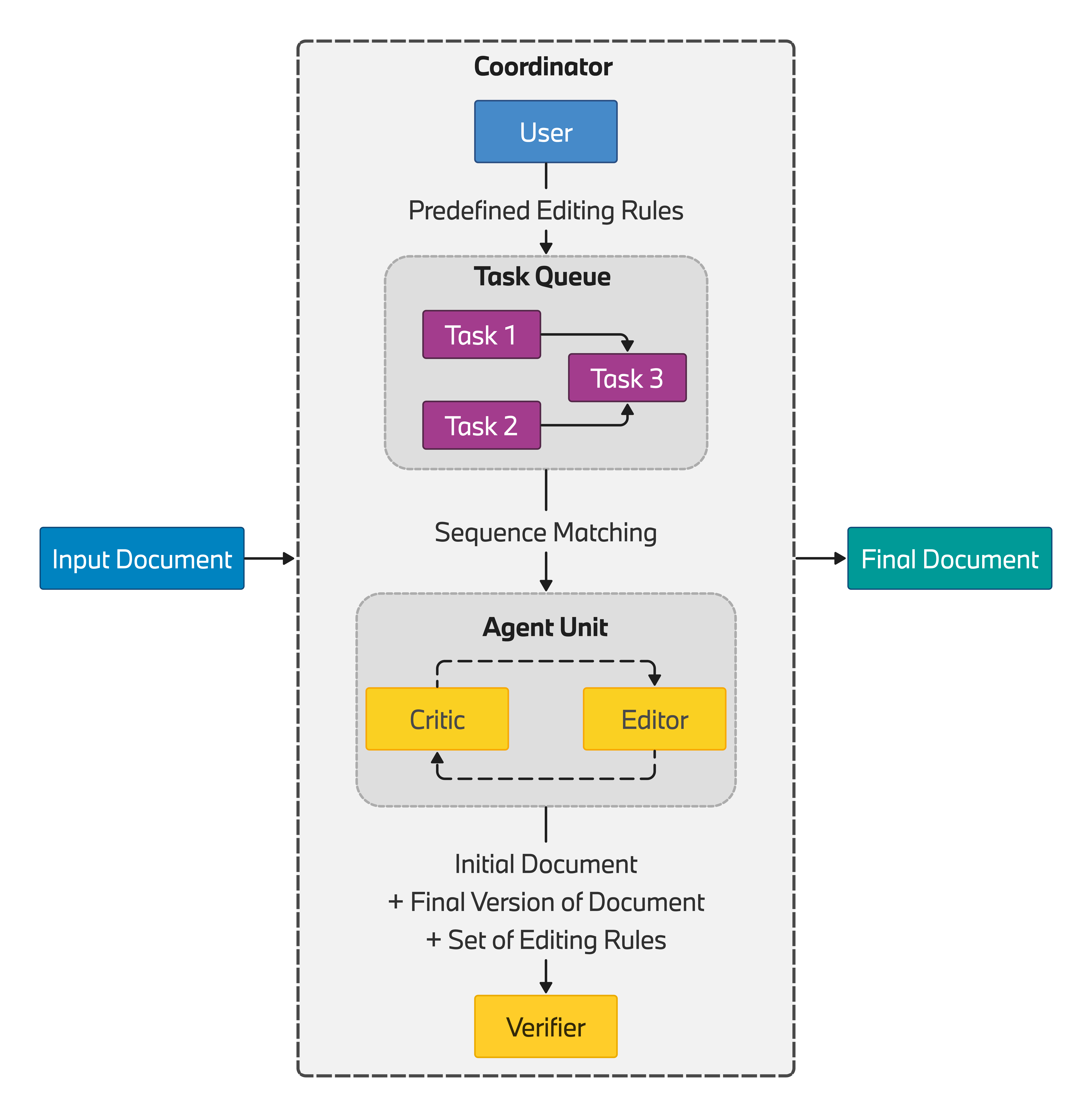}
\caption{Actor/critic workflow for document editing with two agents and
user defined editing rules. The Agent Unit alternates between \textbf{Editor} and
\textbf{Critic} agents via a sequential
matching function.
\label{figure-actor-critic-example}}
\end{figure}

In this example, we consider document editing via two agents, \textbf{Editor} and \textbf{Critic},
following the actor/critic execution model. We do not
utilize a Planner agent, instead we execute a plan that has been predefined as a set of rules for editing the document. The set of rules will
be converted into tasks with a linear dependency to be executed one
after another. The detailed workflow is displayed in Figure 
\ref{figure-actor-critic-example}.

We create an Agent Unit with two agents, \textbf{Editor} tasked with
editing the document according to the given rule, and \textbf{Critic}
which checks if the rule has been fully
applied. The Agent Unit is iterated with the \textbf{Sequence Matcher}
which chooses the next agent based on a provided sequence, which
in our case is Editor and Critic alternating. This sequence is repeated
until the Critic's
message indicates that there are no further edits required. This process is
repeated for the all tasks. The workflow ends with the \textbf{Verifier} receiving
original and final document together with the set of user provided
rules to ensure all rules have been
fully applied to the original document.

\subsection{Coding Tasks - Joint
Collaboration}\label{coding-tasks---joint-collaboration}

\begin{figure}[!ht]
\centering
\includegraphics[width=0.9\textwidth]{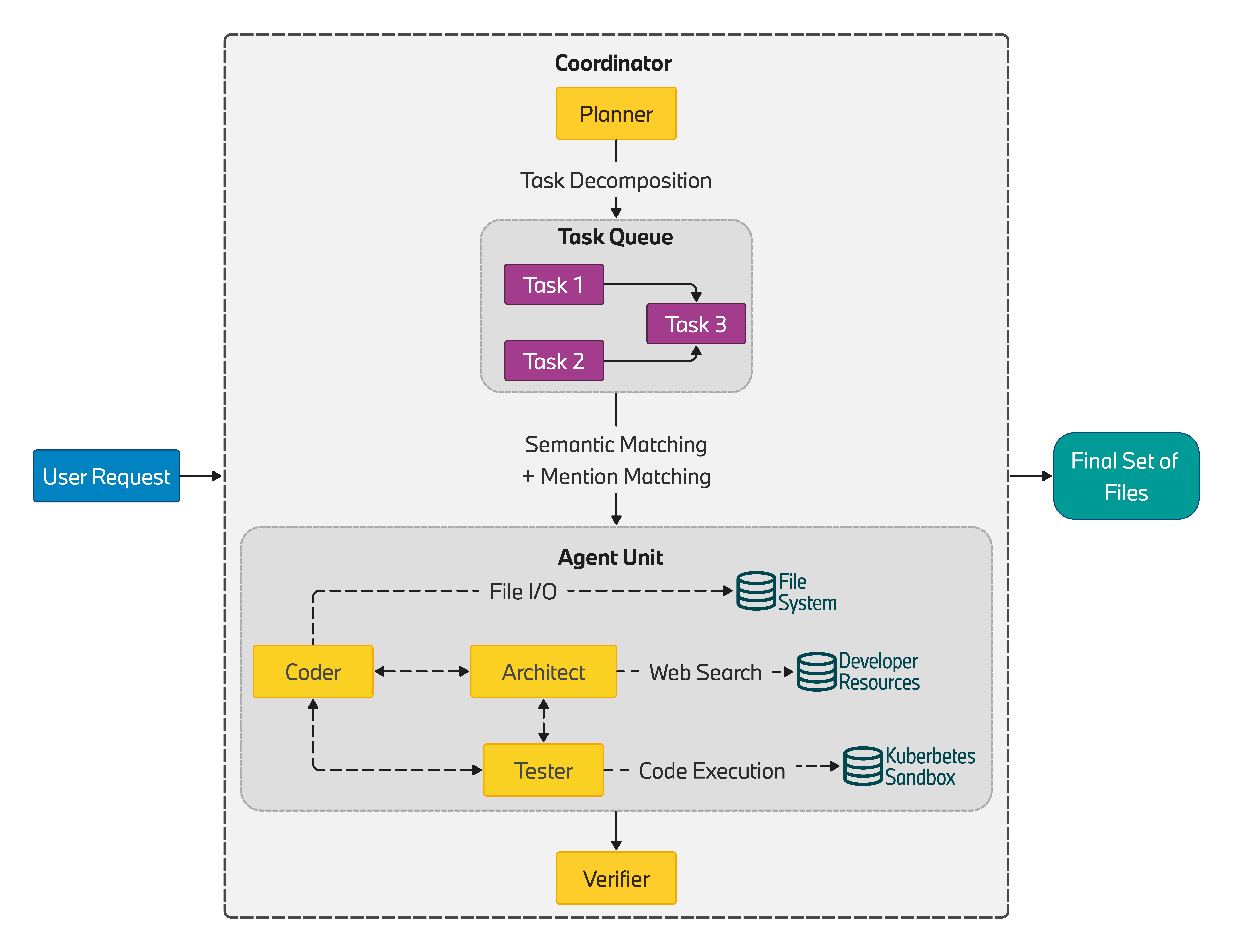}
\caption{Basic workflow for software development with three agents in a
conversational workflow. The Agent Unit consists of \textbf{Coder},
\textbf{Architect} and \textbf{Tester} Agents that can engage in a dialog with each other. The Agent Unit is iterated through a
combination of the Semantic and Mention matching functions with the goal of
selecting the best agent to start and the next agent via explicit message passing. Each Agent has a distinct tool to accomplish its responsibilities.
\label{figure-coding-example}}
\end{figure}

In the final example, we consider a software
development example. This workflow starts with the user request passed to the
\textbf{Planner} Agent to
create a set of tasks that populates the \textbf{Task Queue}. The \textbf{Agent Unit} consists of three
agents with distinct responsibilities: 1) \textbf{Coder} is responsible for
all software engineering tasks using a \textbf{File I/O} tool, 2) \textbf{Architect} creates the architecture for the development using a \textbf{Web Search} tool, and 3) \textbf{Tester} is responsible
for testing the code using a \textbf{Code Execution} tool. All agents are aware of the existence and expertise of the other agents. The Agent Unit is iterated through a combination of the semantic
and mention matching functions which selects the next agent.
\textbf{Semantic Matcher} picks the most appropriate agent for a given
task, \textbf{Mention Matcher} ensures that the conversational strategy
is executed correctly via message passing. Figure \ref{figure-coding-example} illustrates the workflow for this example.
We employ the conversational strategy described in Section 
\ref{multi-agent-workflow}, illustrated with dashed lines between all
agents. The agents work together to produce functioning code that satisfies the code-related task. The workflow ends with the \textbf{Verifier}
Agent that checks the final result against the original user request.

\section{Summary}\label{summary}

In this report, we detail a blueprint 
for a multi-agent engineering framework that aims to address the gaps in existing agent frameworks that may hinder production scale applications \cite{xie2023openagents}. We have presented the main
concepts of our multi-agent engineering framework and
provided example workflows to illustrate the flexibility of our approach. 

To close the gap between existing work and our implementation, we have
considered the following:

\begin{itemize}

\item
  \textbf{Stable Conversational Prompting Strategy} - We have introduced
  the ConvPlanReAct as an approach to bring dialog capability to the ReAct type prompt. We highlight the ability to include all
  stages of the original approach augmented with stages directly relating to
  the multi-agent nature of task execution.
\item
  \textbf{Tools} - We highlight scalable and accurate usage of tools by 
  incorporating the concept of refining the total list available at a given step.
  The refinement mechanisms is coupled with implicit schema for input and output
  of the tool providing ability for accurate usage of underlying functions.
\item
  \textbf{Experiential Learning} - We include the concept of Episodic Memory
  with its direct usage during task execution to retrieve the
  results of previous tasks that are relevant to the current
  one. This retrieval can be scoped to the current plan as well as all previously executed
  plans in the existing application. This enables agents to have access to relevant previous work that has been
  completed.
\item
  \textbf{Human Interaction} - We discuss the inclusion of both intentional and incidental
  human feedback provided back to the agents. In that discussion, we give
  details on how the additional feedback may be provided in the
  proposed ConvPlanReact strategy.
\item
  \textbf{Restart Limitations} - Our approach allows
  for a full restart of any workflow up to the last executed task using
  the information provided in the Episodic Memory. 
  We have introduced a functionality that allows for resuming the workflows that have been paused due to an error, an internal trigger, or an external trigger. 
\end{itemize}

Our framework aims to provide a scalable and highly
configurable environment for multi-agent workflows, enabling stable and production ready agent applications. We have given special consideration to improving multi-agent collaboration, robust tool implementations, extended learning capabilities, incorporating human interactions, and resuming of workflows. These functionalities not only enable higher quality results, they allow us to move beyond experimentation into enterprise settings. 

\bibliographystyle{hunsrtnat}
\bibliography{references}

\end{document}